\documentclass[11pt,a4paper]{article}
\usepackage[utf8]{inputenc}     
\usepackage[T1]{fontenc}        
\usepackage{lmodern}            
\usepackage[british]{babel}
\usepackage{amsmath}
\usepackage{amssymb}
\usepackage{hyperref}
\usepackage{dcolumn}
\usepackage{color}
\usepackage{setspace}           
\usepackage[pdftex]{graphicx}

\usepackage{lmodern}
\usepackage[british]{babel}
\usepackage{graphicx}
\usepackage{grffile}
\usepackage{hyperref}
\usepackage{color}
\usepackage{slashed}
\usepackage{textcomp}
\usepackage{subfigure}
\usepackage{amsmath}
\usepackage{amssymb}
\usepackage{amsfonts}
\usepackage{lscape}
\usepackage{psfrag}
\usepackage{verbatim}
\usepackage{authblk}

\newcommand{\arxiv}[1]{arXiv:\,\href{http://arxiv.org/abs/#1}{{\tt #1}}}

\title{Monte-Carlo simulations of overlap Majorana fermions}
\author[1]{Stefano Piemonte\thanks{stefano.piemonte@ur.de}}
\author[2,3]{Georg Bergner\thanks{georg.bergner@uni-jena.de}}
\author[2]{Camilo L\'opez\thanks{camilo.lopez@uni-jena.de}}
\affil[1]{University of Regensburg, Institute for Theoretical Physics, 
	Universit\"atsstr.~31, D-93040 Regensburg, Germany}
\affil[2]{University of Jena, Institute for Theoretical Physics, 
	Max-Wien-Platz 1, D-07743 Jena, Germany}
\affil[3]{University of M\"unster, Institute for Theoretical Physics, 
	Wilhelm-Klemm-Str.~9, D-48149 M\"unster, Germany}
\begin{document}

\maketitle

\begin{abstract}
Supersymmetric Yang-Mills (SYM) theories in four dimensions exhibit many interesting non-perturbative phenomena that can be studied by means of Monte Carlo lattice simulations. However, the lattice regularization breaks supersymmetry explicitly, and in general a fine tuning of a large number of parameters is required to correctly extrapolate the theory to the continuum limit. From this perspective, it is important to preserve on the lattice as many symmetries of the original continuum action as possible. Chiral symmetry for instance prevents an additive renormalization of the fermion mass. A (modified) version of chiral symmetry can be preserved exactly if the Dirac operator fulfills the Ginsparg-Wilson relation. In this contribution, we present an exploratory non-perturbative study of $\mathcal{N}=1$ supersymmetric Yang-Mills theory using the overlap formalism to preserve chiral symmetry at non-zero lattice spacings. $\mathcal{N}=1$ SYM is an ideal benchmark toward the extension of our studies to more complex supersymmetric theories, as the only parameter to be tuned is the gluino mass. Overlap fermions allow therefore to simulate the theory without fine-tuning. We compare our approach to previous investigations of the same theory, and we present clear evidences for gluino condensation.
\end{abstract}

\section{Introduction}

Supersymmetric Yang-Mills theories (SYM) are promising extensions of the Standard Model (SM) to energies of the order and beyond the TeV scale. However, as the efforts for the experimental discovery of the predicted super-particles are still unsuccessful, the interest to SUSY has been recently motivated by the possibility of understanding in supersymmetric theories many non-perturbative features of strong interactions. Yang-Mills theories exhibit many emergent low-energy phenomena, like confinement and chiral symmetry breaking, that are difficult to understand analytically. The conjecture that there exist supersymmetric models that share the same physical properties if their coupling constant $g$ is exchanged by its inverse $1/g$ opens new perspectives for the analytical understanding of the behavior of Yang-Mills theories. Well-known examples of such connection between different models are the AdS/CFT correspondence and the Seiberg-Witten electromagnetic duality.

The electromagnetic duality has been proposed originally for a certain class of theories in Ref.~\cite{SEI94,SEI94b}, such as SuperQCD and $\mathcal{N}=2$ Supersymmetric Yang-Mills theory. Lattice Monte-Carlo simulations are a useful tool to test non-perturbatively the validity of the Seiberg-Witten duality, in particular when generalizing to $\mathcal{N}=1$ SYM or QCD the proposed mechanisms of confinement, such as the dual superconductor picture. The extension of the electromagnetic duality to more realistic models with less or no supersymmetry requires, in fact, to control the behavior of the theory in the limit, where the mass of the scalar and gluino fields becomes large. However, in this limit SUSY is partially or softly broken and completely different confinement scenarios could be realized. Lattice simulations would be able to investigate a-priori and ab-initio the phase diagram of strong interactions with intact, broken, or partially-broken SUSY.

The presence of many different fermion and scalar fields in the Lagrangian of $\mathcal{N}=2$ SYM and SuperQCD has the consequence to complicate the fine-tuning of lattice simulations. In general, supersymmetry is explicitly broken on the lattice and it can be restored at large distances as an accidental symmetry only after an accurate tuning of the bare couplings and masses of the action. All couplings corresponding to relevant operators compatible with lattice symmetries must be considered, resulting into a number of parameters to tune of $O(10)$, depending on the gauge group and on the number of fields in the original continuum action. $\mathcal{N}=1$ SYM is an exceptional case, since the only relevant parameter to be tuned is the gluino mass. 

The tuning problem can be simplified drastically if a (modified) chiral symmetry is preserved on the lattice by using Ginsparg-Wilson fermions. For instance, the fermion mass would be protected by chiral symmetry, and Yukawa interactions would couple correctly left-handed and right-handed spinors to the corresponding scalar fields. The advantages of chiral fermions might be crucial for simulations of certain supersymmetric theories. Ginsparg-Wilson fermions allow to simulate in principle $\mathcal{N}=1$ SYM without any fine-tuning.

Models with enlarged SUSY can be constructed from simpler supersymmetric theories living in a space-time with extra dimensions. The pure $\mathcal{N}=2$ SYM in four dimensions can be obtained from the compactification of $\mathcal{N}=1$ SYM in six dimensions \cite{BRI77}. Such construction is particularly interesting from the perspective of lattice simulations \cite{MAR97}. If $\mathcal{N}=2$ SYM arises as an effective four-dimensional theory, the scalar mass and the quartic scalar potential would be protected by gauge symmetry. This is due to the fact, that scalars emerge in the small compactification limit from the gauge fields living along the extra dimensions \cite{MAR97,SUZ05}.

In this paper we explore overlap lattice simulations for the pure $\mathcal{N}=1$ SYM in four dimensions. The study of the overlap formulation for $\mathcal{N}=1$ SYM is a crucial step toward the extension of Monte Carlo simulations of more complex supersymmetric theories, a step required to solve the many numerical challenges coming from the Majorana nature of the gluino. In section \ref{N2} we briefly describe the formulation of $\mathcal{N}=1$ SYM, while in section \ref{ov} we review the overlap formalism. In section \ref{tuning} we discuss in details the implementation of the overlap formalism for $\mathcal{N}=1$ SYM in four dimensions and in section \ref{simulations} we summarize the parameters of the lattice simulations. In section \ref{sec:scale} and \ref{sec:gluino} we present the determination of the scale and of the gluino condensate. In section \ref{sec:witten} and \ref{sec:volume} we briefly discuss the expected consequences of volume reduction and of the fermion-boson cancellation of the Witten index on bulk quantities. Finally, before the outlook and the conclusions, in section \ref{sec:spectrum} we present a first determination of the bound state spectrum using Ginsparg-Wilson chiral fermions.

\section{$\mathcal{N}=1$ Supersymmetric Yang-Mills theory}\label{N2}

The simplest Yang-Mills theory with a single conserved supercharge is $\mathcal{N}=1$ SYM, describing the interactions between gluons and their fermion superpartners, the gluinos. The on-shell action reads 
\begin{equation}
 \mathcal{S}_{\mathcal{N}=1\textrm{ SYM}} = \int d^4 x \left\{-\frac{1}{4} F^{\mu\nu}_a F_{\mu\nu}^a + \frac{i}{2} \bar{\lambda}_a \gamma_\mu (\nabla_\mu)_{ab} \lambda_b + \frac{ \theta}{32\pi^2} g^2 F^{\mu\nu}\tilde{F}_{\mu\nu}\right\}\,,
\end{equation}
being $F_{\mu\nu}$ and $\tilde{F}_{\mu\nu}$ the field-strength tensor and its dual. The covariant derivative $(\nabla_\mu)_{ab}$ acts on the gluino fields $\lambda_b$ in the adjoint representation of the gauge group SU($N_c$). The gluino is a Majorana fermion. The angle $\theta$ is set to zero in Monte-Carlo lattice simulations.

Intact supersymmetry implies an exact cancellation between fermion and boson loops in a perturbative asymptotic expansion, allowing an analytic calculation of many quantities and properties that would be out the reach of perturbation theory in non-supersymmetric theories. A remarkable example is the exact $\beta$-function, representing the dependence of the strong coupling constant on the scale. $\mathcal{N}=1$ supersymmetric Yang-Mills theory is asymptotically free and the $\beta$-function is known exactly, including non-perturbative contributions, from instanton calculus \cite{NOV83}. Asymptotic freedom implies that an ultraviolet cut-off, such as $1/a$ of the lattice discretized theory, can be sent to infinity in the limit $g\rightarrow 0^+$ \cite{BER17}. No further tunings are required if chiral symmetry is preserved exactly in the regularized theory, even at energy scales of the order of the ultraviolet cut-off.

\section{Overlap formalism}\label{ov}

The overlap formalism allows to construct a fermion action with an exact (modified) chiral symmetry even at non-zero lattice spacing. The overlap operator is defined as 
\begin{equation}
 D_{\textrm{ov}}(\mu) = \frac{1+\mu}{2}+\frac{1-\mu}{2}\gamma_5\textrm{sign}(D_H)\,,
\end{equation}
where $D_H = \gamma_5 D_W(\kappa)$ is the hermitian Dirac-Wilson operator. The parameter $\mu$ is proportional to the fermion mass, up to a multiplicative renormalization constant.

The mass-less overlap operator fulfills the modified chiral anticommutator 
\begin{equation}
 \{\gamma_5, D_{\textrm{ov}}(0) \} = 2 D_{\textrm{ov}}(0) \gamma_5 D_{\textrm{ov}}(0)\,,
\end{equation}
also known as Ginsparg-Wilson relation \cite{GIS82}. In the mass-less limit, the fermion action $\bar{\psi} D_{\textrm{ov}} \psi$ is invariant under the global chiral transformation proportional to
\begin{eqnarray}
 \delta \psi & \rightarrow & \gamma_5(1-D_{\textrm{ov}}) \psi \\
 \delta \bar{\psi} & \rightarrow & \bar{\psi}(1-D_{\textrm{ov}}) \gamma_5 \,,
\end{eqnarray}
while the variation of the measure of the path integral does reproduce the correct continuum axial anomaly \cite{LUS99}. The factor $(1-D_{\textrm{ov}})$ modifies the definition of the naive continuum chiral rotations, circumventing the Nielsen-Ninomiya theorem \cite{NIE81} and allowing for a chiral symmetric fermion action on the lattice.

The sign function of the hermitian Dirac-Wilson operator can be computed from an eigenvalue decomposition, but for practical numerical calculation it is better expressed as
\begin{equation}
 \textrm{sign}(D_H(\kappa)) = \frac{D_H(\kappa)}{\sqrt{D_H(\kappa) D_H(\kappa)}}\,.
\end{equation}
The inverse square root function is typically approximated by means of polynomial or rational approximations.

Chiral symmetry prevents additive renormalization of the fermion mass, therefore the renormalized mass is proportional to $\mu$. The Dirac-Wilson operator used in the sign function depends on the parameter $\kappa=\frac{1}{8-2\hat{m}}$. The mass $\hat{m}$, not to be confused with the fermion mass $\mu$ of the Overlap operator itself, can in principle take any negative value between $-2$ and zero. 

The fermion propagator in the mass-less case $\mu=0$ is constructed from the inverse of the overlap operator up to a contact term
\begin{equation}
 S_{\textrm{ov}}(0) = (D_{\textrm{ov}}(0)^{-1} - 1)\,,
\end{equation}
such that the propagator anticommutes with $\gamma_5$
\begin{equation}
 \{\gamma_5, S_{\textrm{ov}} \} =  0\,,
\end{equation}
as it follows straight from the Ginsparg-Wilson relation.
It can be shown that quark bilinear operators and interpolators computed from $S_{\textrm{ov}}$ are automatically on-shell and off-shell $O(a)$ improved \cite{CAP99}. The fermion propagator in the mass-less limit can be also rewritten as 
\begin{equation}
 S_{\textrm{ov}}(0) = (1 - D_{\textrm{ov}}(0))D_{\textrm{ov}}(0)^{-1} = (1 - D_{\textrm{ov}}(0))^{1/2}D_{\textrm{ov}}(0)^{-1}(1 - D_{\textrm{ov}})(0)^{1/2} \,,
\end{equation}
that is the propagator of the ``rotated'' fermion fields $\hat{\psi}$
\begin{equation}
 \hat{\psi} = (1 - D_{\textrm{ov}}(0))^{1/2} \psi\,.
\end{equation}
From the latter observation, it is natural to set the massive fermion propagator to be equal to
\begin{equation}
 S_{\textrm{ov}} = (1-D_{\textrm{ov}}(0))D_{\textrm{ov}}^{-1}(\mu) \,.
\end{equation}
Note that the propagator defined with the prescription above is manifestly non-local, contrary to the operator $D_{\textrm{ov}}$, in agreement with the Nielsen-Ninomiya theorem.

\section{Implementation of the overlap gluino algorithm}\label{tuning}

The implementation of the Hybrid Monte Carlo (HMC) algorithm for overlap gluinos has to face several challenges compared to the corresponding QCD simulations. There are two main obstacles. The first is the general problem of topology and fermion zero modes, and the second is the specific problem of the Majorana nature of the gluino field. Several strategies can be applied to control these problems, like the Domain-Wall representation of the Ginsparg-Wilson relation or simulations at fixed topology, which all imply a deviation from the exact chiral symmetric action. We apply instead a polynomial approximation of the sign function as a simple strategy to control the simulations. 

Supersymmetry requires that the gluino mass is exactly zero. Even in the continuum, the zero mass limit is quite peculiar and requires a careful consideration of the order of chiral and infinite volume limit. As a consequence of chiral symmetry, related problems arise with the overlap operator on the lattice. Each lattice configuration featuring a non-trivial topology has at least one zero mode of the overlap operator, such that its determinant and its Boltzmann probabilistic weight are both exactly zero. This simple consequence of preserving chiral symmetry exactly effectively prevents topological charge fluctuations and the generation of instanton-like configurations. However, zero modes and topological objects are also a fundamental feature of the mechanisms responsible for gluino condensation \cite{MOR88,SHI88}. The situation can be effectively described as ``zero over zero'' problem \cite{CRE06}. A regularization is needed for a meaningful definition of observables like the chiral condensate. Like in lattice QCD, this requires a careful consideration of the order of thermodynamic, chiral, and continuum limit. At finite volume and finite lattice spacing, a regulated version of chiral symmetry has to be considered. A related more technical aspect is the divergent HMC fermion force whenever a zero mode is approached leading to a strongly suppressed sampling of topological sectors. The zero modes of the overlap operator can be avoided by a controlled approximation, as the jump of the sign function at the origin is effectively smoothened. We explore the approximation of the sign function by a polynomial of order $N$ to control these problems.

The integration of Majorana fermions leads to the Pfaffian of the operator $D_{\textrm{ov}}$, that is equal to the square root of the determinant.
 The square root of a large sparse matrix cannot be computed exactly, and therefore we are forced to use a second approximation besides the one required to compute the sign function of the overlap operator itself. Similar problems arise in QCD for the strange quark action or for isospin breaking effects in the up and down quark action. In these situations the overlap operator for Dirac fermions can, however, be suitably chirally decomposed so to avoid a twofold approximation \cite{JUC14}, but such decomposition does not extend directly to Majorana fermions.
In our overlap simulations we proceed in analogy with the Wilson case, and we compute the fourth root of the positive-definite square of the hermitian operator $\gamma_5 D_{\textrm{ov}}$
\begin{equation}
\textrm{Pf}(D_{\textrm{ov}}) = \sqrt{\det{(D_{\textrm{ov}})}} = \sqrt{\det{(\gamma_5)}\det{(D_{\textrm{ov}})}} = \sqrt[4]{\det(\gamma_5 D_{\textrm{ov}})^2}\,.
\end{equation}
A rigorous treatment of Majorana fermions on the lattice leading to the above equations in the overlap formalism has been presented in Ref.~\cite{NAR97}. The fourth root of the square hermitian operator is approximated by a rational fraction of two polynomials (RHMC). Pseudofermion fields $\phi$ are introduced at the beginning of trajectory leading to the effective action 
\begin{equation}
S_f = \phi^\dag ((\gamma_5 D_{\textrm{ov}})^2)^{-\frac{1}{4}}\phi\,,
\end{equation}
that is integrated numerically. Contrary to Wilson fermion, we will demonstrate that the approximated overlap operator does not lead to a sign problem.
The second main reason for our approach based on polynomial approximations is a stable precision of the RHMC algorithm. If we would consider a rational instead of the polynomial approximation of the overlap operator, an additional inner inverter inside the outer rational approximation of the RHMC would be required and a stable algorithm is in general not guaranteed. 

Several possibilities to overcome the challenge of the HMC algorithm due to the non-differentiability of the sign function have been discussed in the literature \cite{FOD03,FOD04,FUK06,MAT06,WEN06,DEG06,DEG08,CUN08}. Each time a zero mode of the hermitian Dirac-Wilson operator $D_H$ crosses the origin, there is a jump in the topological charge and a discontinuity in the fermion force. A solution is to monitor the changes of the lowest eigenvalues of $D_H$ during the HMC trajectory and perform a refraction/reflection step \cite{FOD03,FOD04}. The Arnoldi algorithm can provide an accurate computation of the smallest eigenvalues of $D_H$, but the numerical cost is quite huge. Another possibility is to perform simulations at fixed topological charge, thereby forbidding discontinuities of the force \cite{FUK06}. Fixed topological charge simulations might however fail to address important physical features of the vacuum structure of SYM theories where topological objects are assumed to play a relevant role, such as gluino condensation \cite{COH84}.

Our approach is different from the preliminary studies with overlap gluinos presented in Ref.~\cite{JLQ11}, where an additional term has been included to the action to generate a spectral gap in the operator $D_H$ and to fix the simulations to the topological sector zero. This is not necessary with a polynomial approximation at finite $N$. Our approach is instead similar to the simulations with Domain-Wall fermions, with the order of the polynomial approximation $N$ replaced by the lattice extend $L_5$ in the fifth dimension. In both cases, the Ginsparg-Wilson relation is only approximately fulfilled, and exact chiral symmetry is recovered in the limit $N\rightarrow \infty$ or $L_5 \rightarrow \infty$ \cite{GIE08,KAP00,WEN06}. Quantities like the gluino condensate are computed after considering the infinite volume limit first, then the extrapolation $N \rightarrow \infty$, and finally the continuum limit. For simplicity, we use chiral limit in the following already for the limit $N\rightarrow \infty$, which always assumes that $N$ is still small enough to keep the finite volume effects and lattice artefacts under control.

As already explained the smoothing of the sign problem by a polynomial approximation reduces the problem of near zero eigenvalues, which can be considered as a controlled way to relax the Ginsparg-Wilson relation. A polynomial of order $O(100)$ might provide a good approximation of the sign function while keeping differentiability of the HMC trajectory. The error of the polynomial approximation could be corrected afterwards by means of reweighting \cite{DEG08}. Depending on the parameters of the simulations, in particular the fermion mass and the volume of the lattice, there will be an optimal tuning as a compromise between the computational cost and the effectiveness of the polynomial approximation. We are going to explore this possibility in the next sections. 

\subsection{Simulations with overlap gluinos}\label{simulations}

\begin{figure}
\centering
 \subfigure{\includegraphics[width=.49\textwidth]{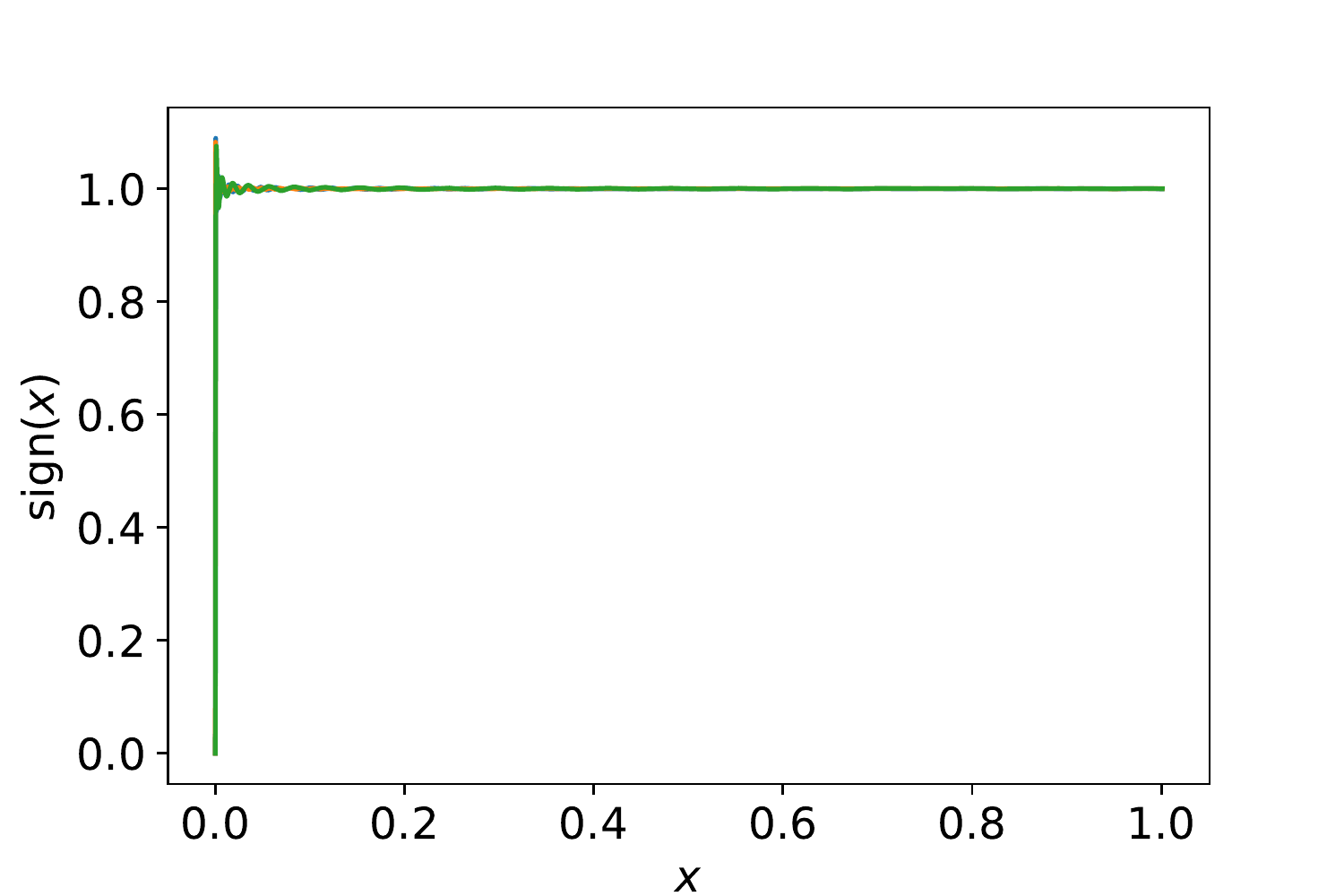}\label{sign_approximation_full}}
 \subfigure{\includegraphics[width=.49\textwidth]{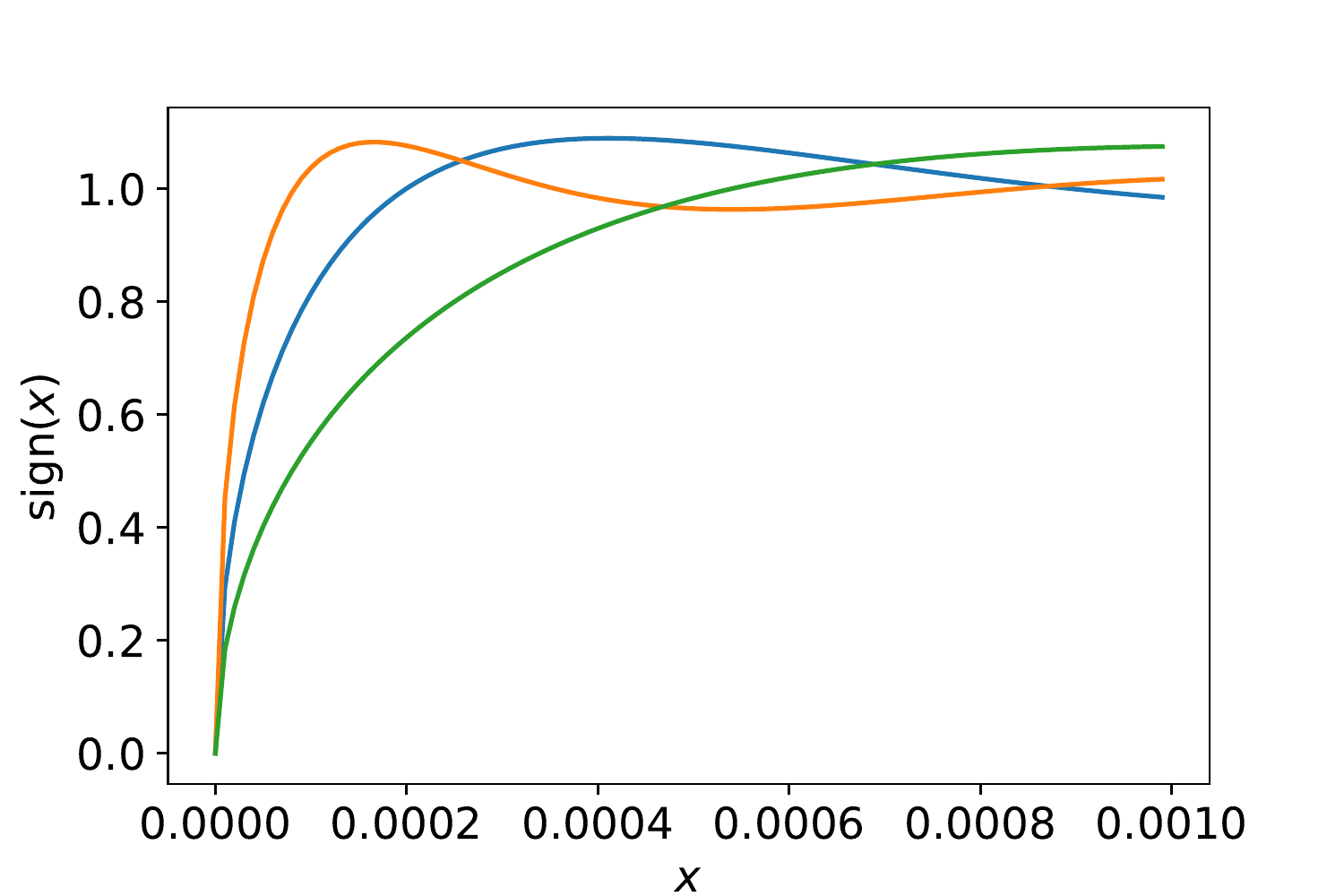}\label{sign_approximation}}
 \caption{Polynomial approximation of the sign function for $n=160$ (green), $n=250$ (blue) and $n=400$ (orange). The inverse square root function is approximated by means of quadratically optimized in the space of the orthogonal polynomials and the sign function is simply computed as $\textrm{sign}(x) = x / \sqrt{x^2}$.}
\end{figure}

Simulations with overlap fermions are significantly more expensive compared those with Wilson fermions. The calculation of the force and of the inverse of the overlap Dirac operator requires at each iteration the evaluation of the approximation of the sign function. Our main studies to check the validity of the proposed algorithm have been done for the gauge group SU(2) on a $8^4$, $12^3$ and $16^3\times 32$ lattice with $\mu=0.0$ and $\beta=1.6$. The complete summary table of all volumes and their corresponding polynomial approximations can be found in App.~\ref{sec:summary}. The code used for the simulations presented in this contribution is publicly available on GitHub \cite{PIE20}. It is a generic code for simulations of Yang-Mills theories for arbitrary number of colors and fermions in adjoint or fundamental representation.

The gauge action is discretized using a combination of $1\times1$ and $2\times1$ Wilson loops, employing the so-called tree-level Symanzik action. The sign function has been approximated using three different polynomials of order $N=160$, $N=250$ and $N=400$, following the strategies and the algorithms presented in Ref.~\cite{MON97}. The approximations are shown in the Fig.~\ref{sign_approximation}. Note that the discontinuity around the origin is smoothed by the polynomial approximation itself. The links for the overlap operator are one-level stout-smeared with smearing parameter $\rho=0.15$. While the hopping parameter of the Dirac-Wilson approximation could in principle be chosen freely in the interval $0.125<\kappa<0.25$, a certain tuning is required such that zero modes of the mass-less overlap operator do appear. We have set $\kappa=0.2$, and we will show that zero modes are present in our simulations.

The rational approximation used for the computation of the force can be chosen to be in principle different from the one used for the metropolis/heatbath steps. We approximate the fourth root of the Dirac operator by two rational approximations with four and eleven fractions, used in combination with multiple time-scales integrators. The rational approximations used for the Metropolis step require instead a higher accuracy and are the sum of 37 fractions. The spectral intervals used to construct the rational approximations are chosen to fit all eigenvalues of the overlap operator, including the lowest modes.

The boundary conditions imposed on fermion fields are relevant when dealing with supersymmetry simulated on a small box. As the difference in thermal statistics between bosons and fermions breaks supersymmetry at non-zero temperature, we impose periodic boundary conditions to all fields in all directions.

\section{Scale setting and topological sampling}\label{sec:scale}

\begin{figure}
\centering
 \subfigure[$w_0/a$]{\includegraphics[width=.49\textwidth]{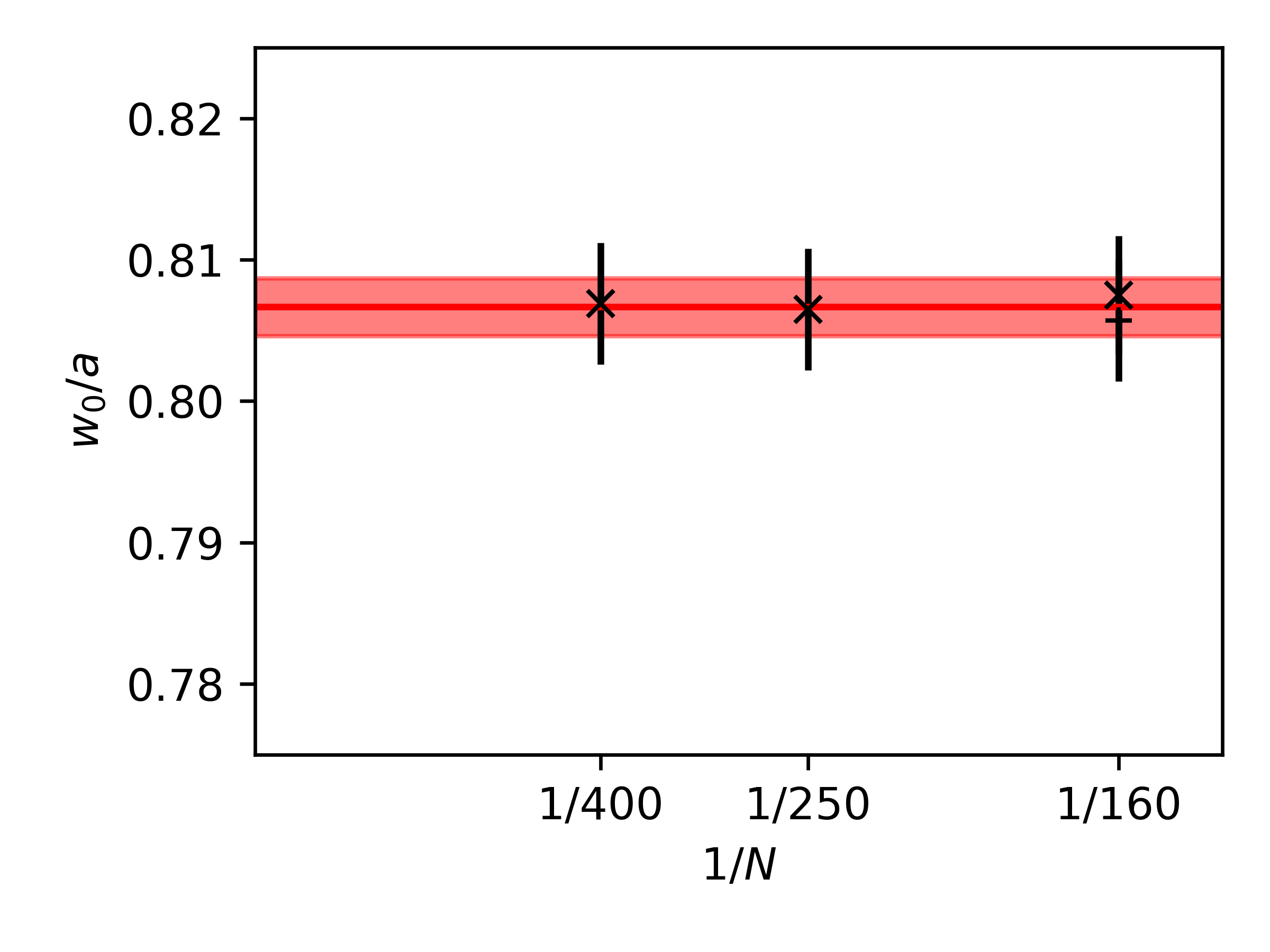}\label{w0_0000m_1600b_2000k}}
 \subfigure[Topological charge]{\includegraphics[width=.49\textwidth]{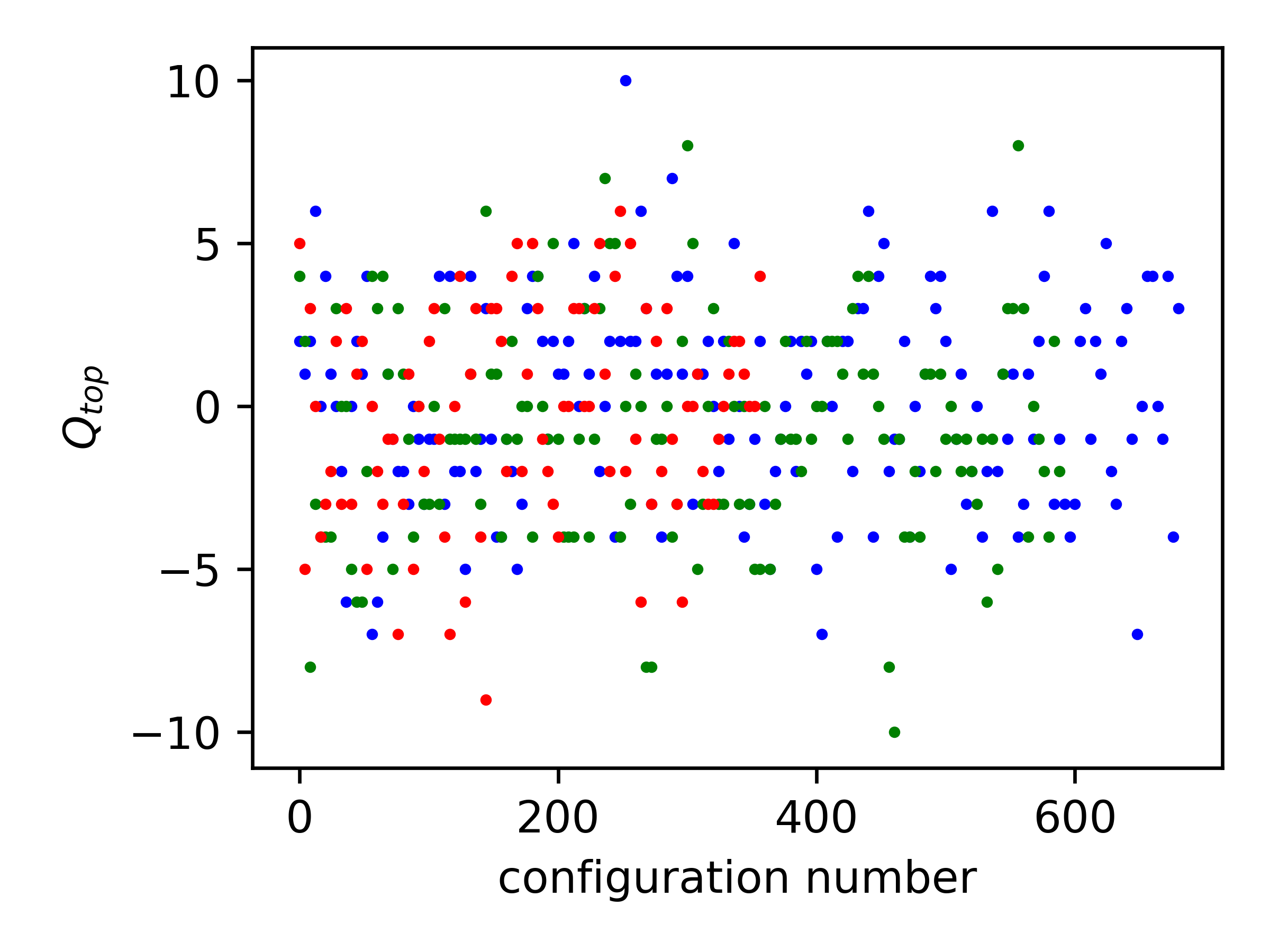}\label{topological_charge_0000m_1600b_2000k}}
 \caption{a) Scale $w_0/a$ as function of the order of the polynomial approximation of the sign function for the volume $12^4$ ($\times$ symbols) and for $N=160$ and a volume $16^3\times 32$ ($+$ symbol). The extrapolation to the chiral limit is shown by the red line with the error given by the red band.  b) Monte Carlo history of the topological charge $Q_{\textrm{top}}$ on a volume $12^4$ for $N=160$ (blue), $N=250$ (green), $N=400$ (red), excluding thermalization.}
\end{figure}

The Lagrangian of $\mathcal{N}=1$ SYM in four dimensions does not depend on any dimensional coupling. The lattice spacing $a$ is a function of the bare gauge coupling $g$ and the scale is dynamically generated by the trace anomaly through dimensional transmutation. The lattice spacing $a$ is expressed in terms of an infrared observable known with good accuracy, such as the Wilson flow scale $w_0/a$, defined by the flow time $\tau$, where the condition 
\begin{equation}
\tau \frac{d}{d \tau} (\tau^2 \langle E(\tau)\rangle ) = 0.3
\end{equation}
is fulfilled. We integrate the flow defined from the Wilson gauge action and the expectation value of the energy density $\langle E(\tau)\rangle$  is measured from its clover antisymmetric definition. For further details, see \cite{BER5}. All masses and low energy constants measured in dimensionless units of $w_0/a$ are expected to scale to the continuum limit with corrections of the order $O(a^2)$. The scale $w_0$ is presented in Fig.~\ref{w0_0000m_1600b_2000k}, and it is basically independent of the order $N$ of the polynomial used for the approximation of the sign function and free from finite size effects already at a volume $12^4$. Therefore, we extrapolate $w_0/a$ to the chiral point ($N\rightarrow\infty$) using a simple constant function, obtaining $w_0/a = 0.8067(20)$.

The chirally extrapolated value of $w_0/a$ yields to a rather coarse lattice spacing compared to the previous SU(2) $\mathcal{N}=1$ SYM simulations of our collaboration using unimproved Wilson fermions. At the current stage of the project we are in particular interested in the effects of the improved chiral symmetry at rather coarse lattice spacings. Future exploratory investigations of rather complex extended supersymmetric theories with scalar fields are difficult at fine lattice spacings and we want to explore the prospects of the presented approach for these studies. Large scale simulations of complex supersymmetric models including scalars are beyond the reach of current computational resources, and it is important to keep the lattice volume as small as possible while being able at the same time of preserving the main features of the continuum theory being simulated.

Simulations at finer lattice spacing with supersymmetry preserving periodic boundary conditions might result into topological freezing. In our current setup, the topological charge is fluctuating between different topological sectors, see Fig.~\ref{topological_charge_0000m_1600b_2000k}. We measure the topological charge from its field theory definition as the discretization of $\tilde{F}_{\mu\nu} F^{\mu\nu}$ measured at the flow time $t=2.0$, rounded in order to be an integer. In principle, a proper definition of the topological charge is given by the chirality of the zero modes of the overlap operator. We plan to compare the two definitions in the near future. In the present calculations, we use the flowed $\tilde{F}_{\mu\nu} F^{\mu\nu}$ as a simple measure of the autocorrelation of our Monte Carlo chains, and we postpone to a forthcoming publication the study of the index theorem in relation to the chiral Ward identities.

\section{Gluino condensation}\label{sec:gluino}

\begin{figure}
\centering
 \subfigure[Chiral condensate $a^3\Sigma$]{\includegraphics[width=.49\textwidth]{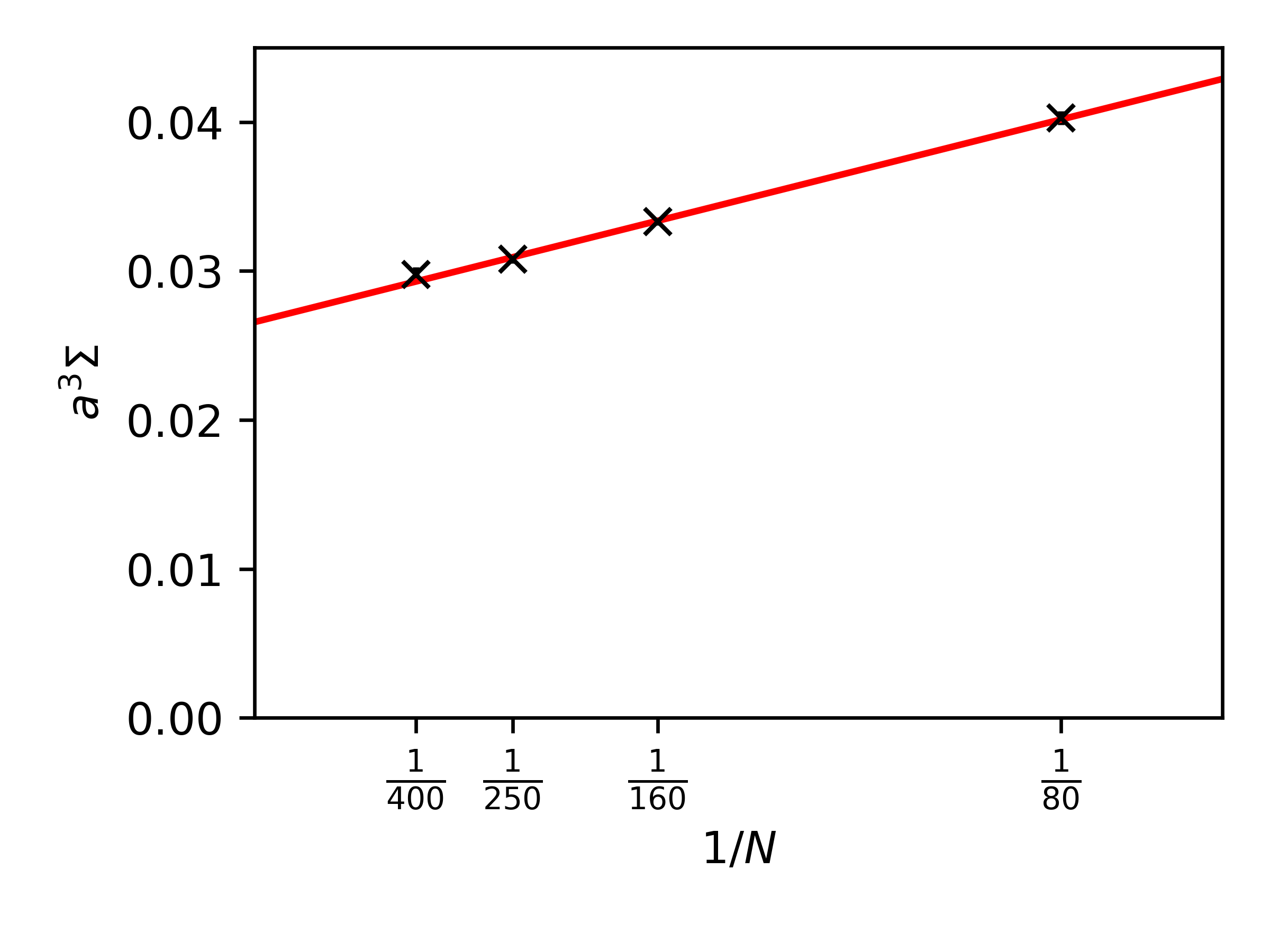}\label{chiral_condensate_0000m_1600b_2000k}}
 \subfigure[Volume dependence of $a^3\Sigma$ for $N=250$]{\includegraphics[width=.49\textwidth]{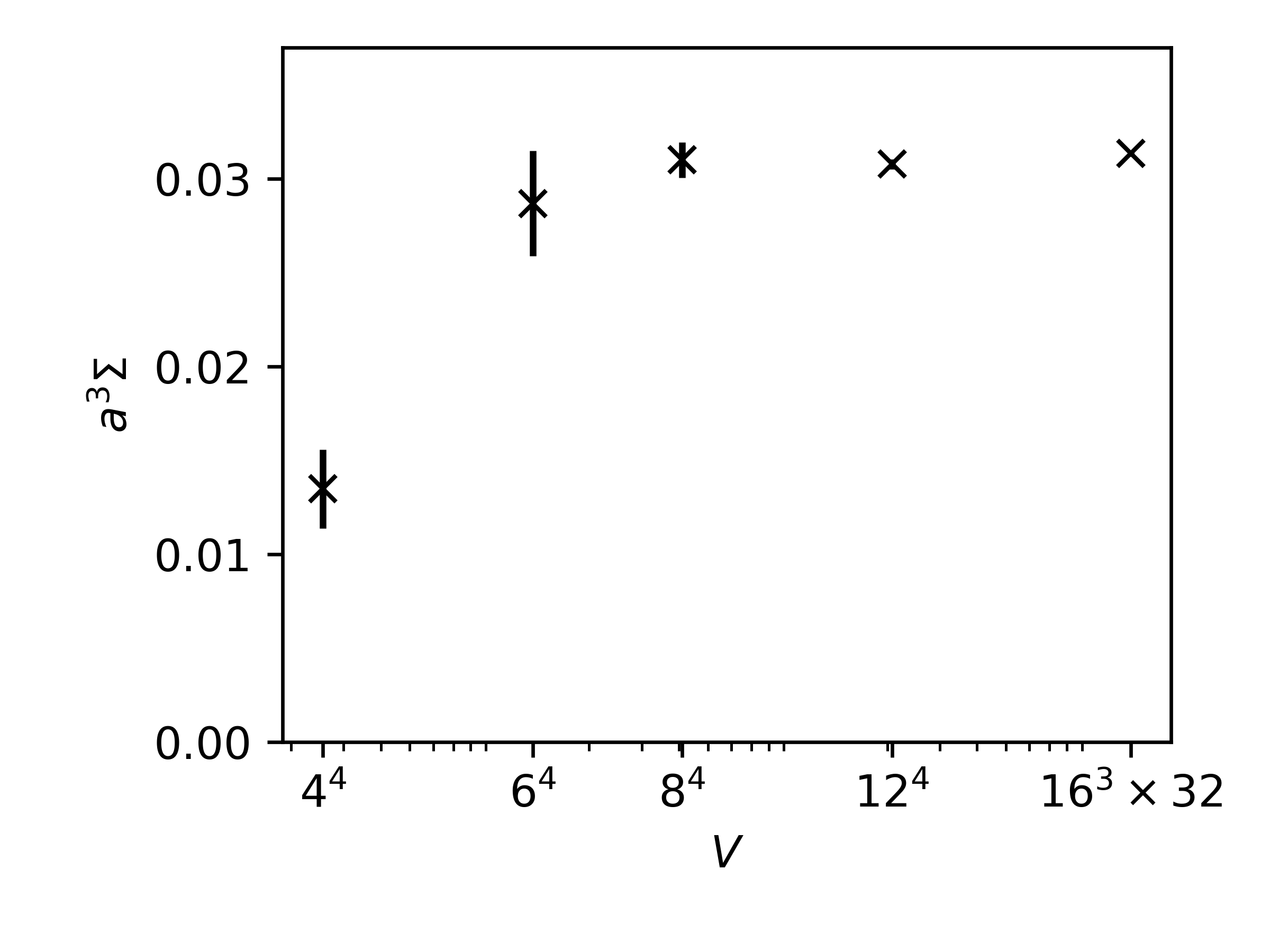}\label{volume_chiral_condensate_250_0000m_1600b_2000k}}
 \caption{a) The extrapolation to the chiral limit of the gluino condensate $a^3\Sigma$ as a function of $1/N$ for the volume $12^4$.  b) Volume dependence of $a^3\Sigma$ as a function of the volume $V$ for the order of polynomial of the sign function $N=250$.}
\end{figure}

Simulations with overlap gluinos open new perspectives beyond the solution to problem of the tuning of supersymmetric Yang-Mills theories to the continuum limit. In particular, the vacuum structure of $\mathcal{N}=1$ SYM might be accessible only with an exact chiral symmetric fermion action. In fact, although $U_A(1)$ axial symmetry of the form 
\begin{equation}
 \lambda \rightarrow \exp{(i \theta \gamma_5)} \lambda
\end{equation}
is broken by the anomaly, the theory has still a remnant discrete $Z_{2N_c}$ fermion symmetry for angles $\theta = 2 \pi \frac{n}{2N_c}$. At zero temperature, this symmetry is spontaneously broken down to $Z_2$ by a non-vanishing expectation value of the gluino condensate \cite{SHI88,MOR88}. The number of vacuum states at zero temperature is $N_c$, equal to the Witten index of $\mathcal{N}=1$ SYM. In particular, if the gauge group is SU$(2)$, the coexistence of two different phases is expected. If fermion doublers are removed from the physical spectrum of the naive fermion action by breaking chiral symmetry with the addition of the Wilson term, it is impossible to directly measure the gluino condensate due to the additive renormalization of the vacuum expectation value of $\langle \bar{\lambda}\lambda\rangle$. Evidences of chiral symmetry breaking can be given only by studying the behavior of the theory at non-zero temperatures or by using the gradient flow as an additional regularization scheme besides the lattice \cite{BER19,BER14a}.

In the overlap formalism the gluino condensate $\Sigma$ is free from additive renormalization. The bare gluino condensate is defined as the expectation value of the fermion bilinear
\begin{equation}\label{sigma}
 \Sigma = \langle \bar{\lambda}(1-D_{\textrm{ov}})\lambda \rangle = \langle \textrm{Tr}((1-D_{\textrm{ov}})D_{\textrm{ov}}(\mu)^{-1})\rangle \,,
\end{equation}
the trace is computed using random vectors and stochastic techniques. Note that the gluino condensate is equivalent to the derivative of the partition function with respect to the mass parameter $\mu$. The gluino condensate $\Sigma$ defined by Eq.~\ref{sigma} resembles the continuum $\langle \bar{\lambda}\lambda\rangle$. In the following discussions of this section, in order to avoid confusions, we will denote the chiral condensate by $\Sigma$ instead of $\langle \bar{\lambda}\lambda\rangle$.

The bare gluino condensate is shown for four different polynomial approximations in Fig.~\ref{chiral_condensate_0000m_1600b_2000k}, and for five different lattice volumes for $N=250$ in Fig.~\ref{volume_chiral_condensate_250_0000m_1600b_2000k}. The gluino condensate exhibits a linear behavior as a function of $1/N$. The extrapolation to the chiral point yields to the bare condensate $a^3 \Sigma = 0.02657(37)$, with a $\chi^2/$dof=0.7. The non-zero value of $\Sigma$ is a direct confirmation of the spontaneous breaking of chiral symmetry, being the gluino condensate proportional to its renormalized value only up to a multiplicative renormalization constant. The chiral condensate is almost independent on the lattice volume up to the small volume $4^4$. 
\begin{figure}
\centering
 \includegraphics[width=.49\textwidth]{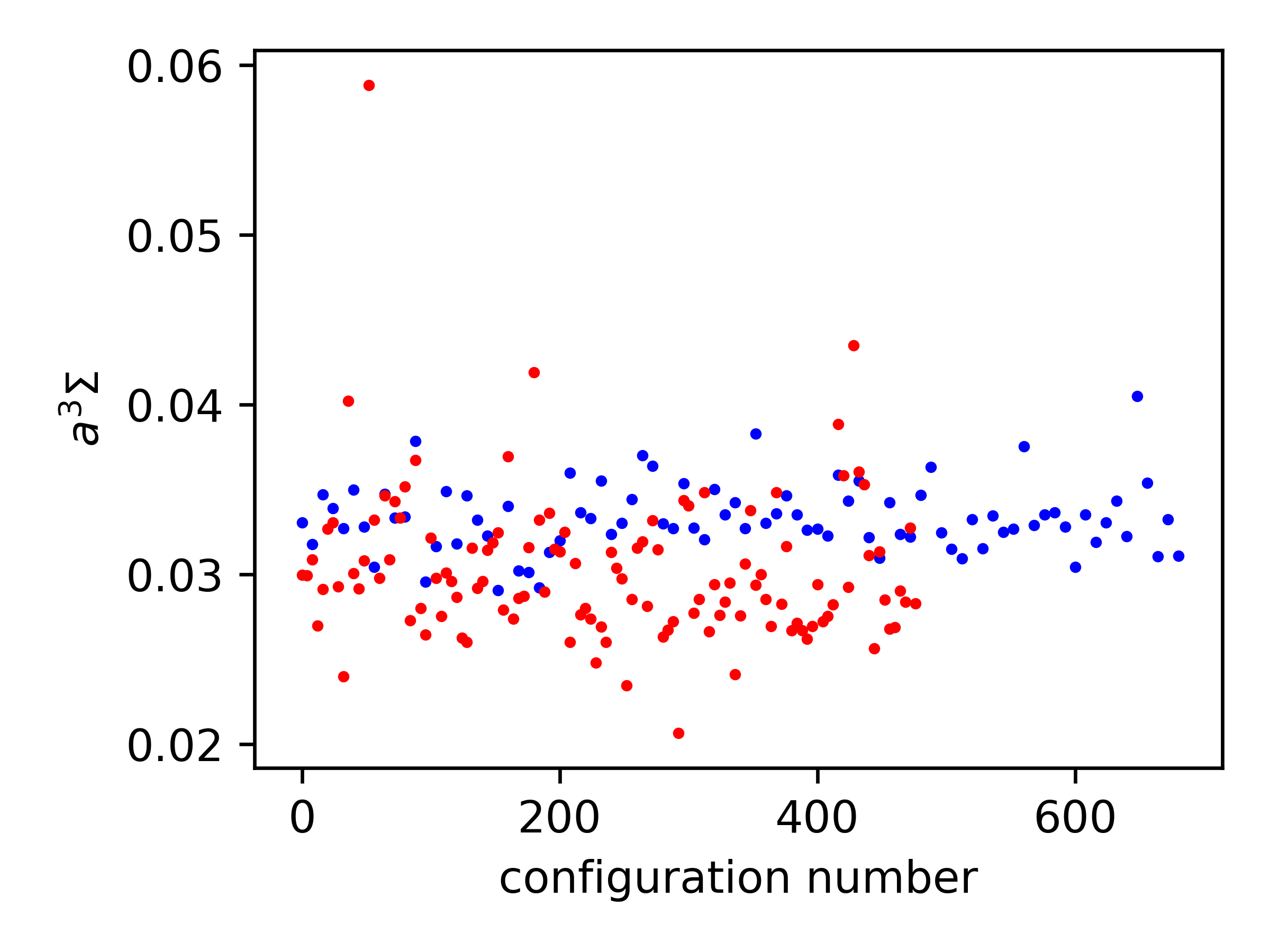}
 \caption{Monte Carlo history of the gluino condensate on a lattice of volume $12^4$ for $N=160$ (blue) and $N=400$ (red).}\label{chiral_condensate_160_vs_400_0000m_1600b_2000k}
\end{figure}

The striking feature of the chiral condensate is its smooth dependence on the order of the polynomial used to approximate the sign function, meaning that our approach has been able to capture the chiral properties of $\mathcal{N}=1$ SYM while regularizing the zero modes of the overlap operator. Looking to the Monte Carlo history of $\Sigma$ in Fig.~\ref{chiral_condensate_160_vs_400_0000m_1600b_2000k}, the mild dependence of the chiral condensate on $1/N$ is a result of two competing effects. On the one side, configurations with small eigenvalues of the overlap operator are rarely generated going closer to the chiral limit $N \rightarrow \infty$, as their Boltzmann weight becomes smaller and smaller. On the other side, once a configuration with small eigenvalues is generated, its individual gluino condensate $\Sigma$ is very large. After taking the ensemble average, the $N=160$ simulation has a very similar chiral condensate to the one with $N=400$, resulting from the average of a large number of configurations with a smaller $\Sigma$ and few of them showing clear upward spikes from the bulk distribution. This phenomenon is effectively the regularization of the ``zero over zero'' problem given by mass-less chiral fermions on the lattice. The non-perturbative dynamics of the theory dictates a non-vanishing expectation value of the order parameter for chiral symmetry breaking, including its precise value, despite the na\"ive reasoning that zero-modes cannot be generated in a Monte Carlo simulations and therefore gluino condensation should not occur. This also implies that there is an upper bound for $N$ at a given volume and lattice spacing to keep the error under control.

The gluino condensate is strictly positive in all our configurations. In the chiral limit, the second phase where the gluino condensate is negative can be reached simply after performing a chiral rotation, in the same way in which all vacuum expectation values of the Polyakov loop in the deconfined phase can be reached by a center symmetry transformation of the temporal links in a given timeslice. The interface tension is however an interesting thermodynamic quantity that is difficult to determine in the current setup, as tunneling during a Monte Carlo simulation of the chiral condensate from positive to negative value is impossible, even on very small volumes. It would be interesting to study the dynamics of the domain-wall interpolating between such two phases, a possible solution to achieve this goal could be given by reweighting combined with a multi-canonical approach \cite{CRE06}.
 
\subsection{Gluino condensate and spectrum of the overlap operator}

\begin{figure}
\centering
 \includegraphics[width=.69\textwidth]{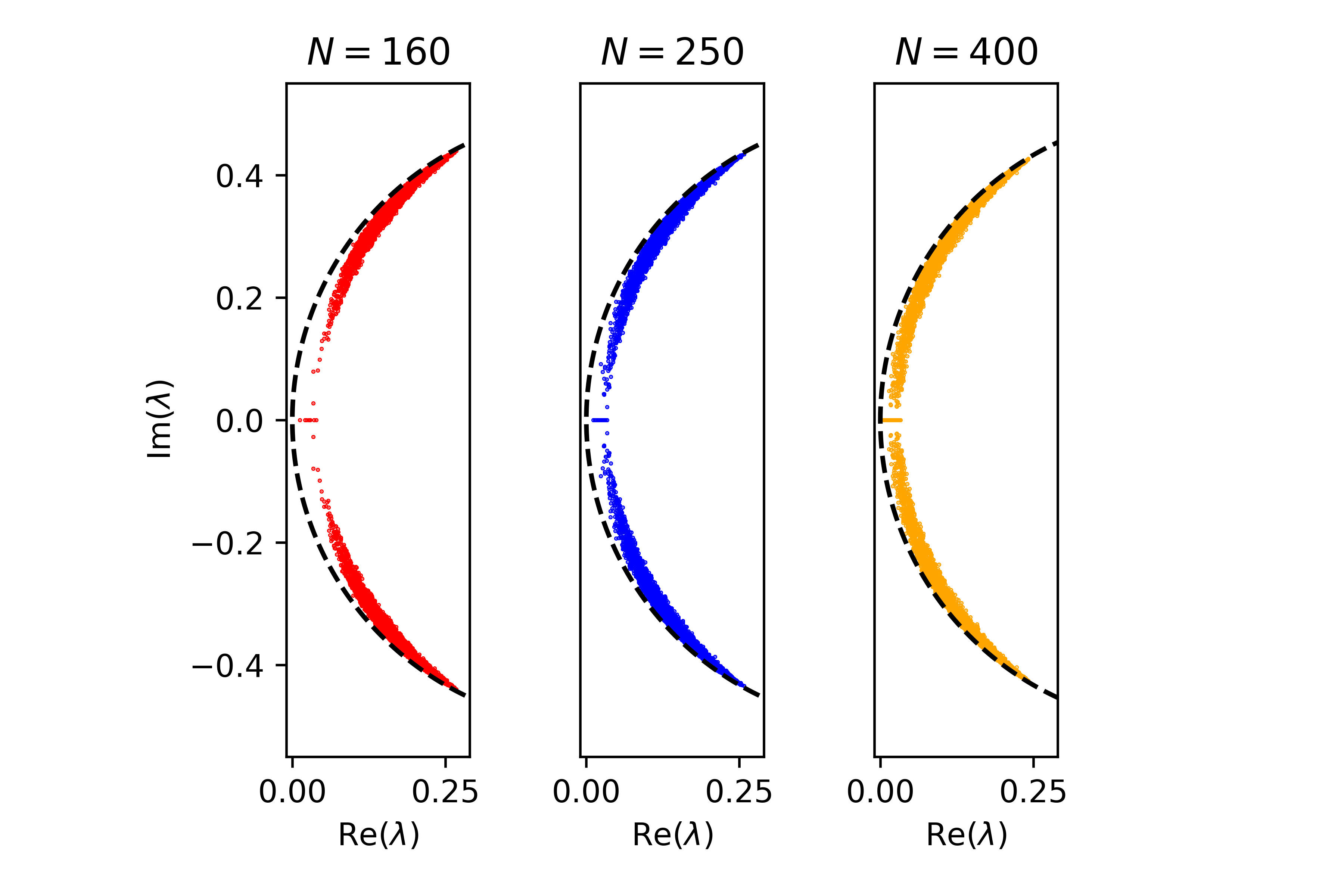}
 \caption{Lowest part of the spectrum of the overlap operator as obtained from an ensemble of configurations on a lattice of volume $8^4$.}\label{full_circle_overlap}
\end{figure}

The dominant contribution to the gluino condensate comes from the eigenvalues of the overlap operator near the origin, and it is crucial to confirm our results by studying the lowest part of the spectrum of $D_{\textrm{ov}}$. 
The eigenvalues of the mass-less overlap operator lie on the complex plane, on a circle of radius $1/2$ centered on the real axis at $x=1/2$; in fact, the matrix $2(D_{\textrm{ov}} - 1)$ is unitary if $\mu=0$, as it follows from the relation
\begin{equation}
 (D_{\textrm{ov}} - 1)^\dag (D_{\textrm{ov}} - 1) = \frac{1}{4} \textrm{sign}(D_H)^2 = \frac{1}{4}\,.
\end{equation}
The spectrum of the smallest eigenvalues from an ensemble of $8^4$ lattice configurations is presented in Fig.~\ref{full_circle_overlap}. As a consequence of the polynomial approximation of the sign function, the eigenvalues are not lying exactly around a circle, but they are somewhat scattered towards its inside. Real eigenvalues close to the origin, of topological nature, are generated for all three polynomial approximations, meaning that the hopping parameter $\kappa$ of the Dirac-Wilson operator used in the definition of the overlap fermion action is correctly tuned. Further, we see that the eigenvalues are bent toward the inward of the overlap circle, and none of them appears on the negative side of the real axis. Therefore, the sign of the Pfaffian is positive on all configurations we have generated, and we have clear numerical evidences that the polynomial approximation of the sign function is able to regularize the RHMC simulations while solving the sign problem. In agreement with the understanding of chiral symmetry breaking from the Banks-Casher relation, we observe that the spectrum of the overlap operator becomes more dense around the origin and is closing the circle in the complex plane in the limit $N\rightarrow\infty$. As the (quenched) mass parameter of the overlap operator would cross the origin going from positive to negative values, the full eigenvalue density around the origin would be responsible for the discontinuity giving rise to the non-zero gluino condensate.

\subsection{Comparison to previous determinations of the gluino condensate}

A full comparison between different determinations of the gluino condensate requires to set a common scale and a common scheme to determine the multiplicative renormalization constants $Z_S$. Previous studies with overlap fermions used the scale $r_0/a$, coming from the static interquark potential, instead of $w_0/a$. This difference is not an obstacle as we have been able to determine the continuum limit extrapolation of the ratio $r_0/w_0=2.21(12)$ in Ref.~\cite{BER5}. After converting all previous results to the scale $w_0$, we have for the bare condensate
\begin{eqnarray*}
w_0^3 \Sigma & = & 0.01394(2)~~~~\textrm{this study,}\\
w_0^3 \Sigma & = & 0.0118(23) ~~~~\textrm{Domain-Wall fermions of Ref. \cite{GIE08},} \\
w_0^3 \Sigma & = & 0.01214(1) ~~~~\textrm{Wilson fermions + gradient flow of Ref. \cite{BER19},}\\
w_0^3 \Sigma & = & 0.058(15)~~~~~~\textrm{Overlap fermions + RMT of Ref. \cite{JLQ11}.}
\end{eqnarray*}
For the Domain-Wall estimation of Ref~\cite{GIE08}, we consider the chiral condensate from $\beta=2.4$ and the $r_0/a$ scale extrapolated to the chiral point as a function of $m_{\textrm{res}}$ from the volume $16^3 \times 32$. Apart from the results of Ref.~\cite{JLQ11} using the overlap formalism but measuring $\Sigma$ from the eigenvalue density and Random Matrix Theory (RMT), all other values of the gluino condensate are close to each other, confirming their equality up to an $O(1)$ renormalization constant. A complete comparison to test the full agreement between the different approaches will be possible only after the multiplicative renormalization constant $Z_S$ is computed in the same scheme at the same scale, and especially only after that $w_0^3 \Sigma$ is extrapolated to the continuum limit, requirements that are currently impossible to meet.

In Ref.~\cite{HIS97}, the authors argued that the renormalization group invariant (RGI) gluino mass $m_{\textrm{RGI}}$ is given by the combination
\begin{equation}
m_{\textrm{RGI}} = \frac{\beta(\alpha)}{\alpha} m_{\tilde{g}}
\end{equation}
where $\beta(\alpha)$ is the well-known NSVZ $\beta$-function. The above relation is valid to all orders in the coupling constant in a scheme that is able to preserve the holomorphic properties of supersymmetry, otherwise it is expected to hold only up to two loops in perturbation theory. Since the condensate is a derivative with respect to the gluino mass $m_{\tilde{g}}$, the RGI value of the condensate can be expressed as
\begin{equation}
w_0^3 \Sigma^{\textrm{RGI}} =\frac{\alpha}{\beta(\alpha)} w_0^3 \Sigma\,.
\end{equation}
This formula provides a simple perturbative estimation independent of the scheme and the scale in the limit $\alpha \rightarrow 0$. The strong coupling constant $\alpha = g^2/(4\pi)$ can be set to be equal to the bare lattice value in the case of the gluino condensate computed straight from the derivative of the logarithm of the lattice partition function. The gradient-flow chiral condensate would require instead a study of the RGI scaling with respect to the gradient flow coupling. The incompatibility of the Overlap and Domain-Wall RGI gluino condensates
\begin{eqnarray*}
w_0^3 \Sigma^{\textrm{RGI}} & = & 0.002827(4)~~\textrm{this study,}\\
w_0^3 \Sigma^{\textrm{RGI}} & = & 0.0016(3) ~~~~~\textrm{Domain-Wall fermions of Ref. \cite{GIE08},} 
\end{eqnarray*}
is most likely due to the missing extrapolation to the continuum limit and due to the bad convergence of perturbation theory in the lattice regularization scheme.

We finally discuss in more detail the comparison of our polynomial approximation and the results from Domain-Wall fermions in Ref. \cite{GIE08}. As explained in the previous section, the limit of large $N$ should be compared to the limit of large $L_5$, the extend in the fifth dimension of the Domain-Wall solution. At $\beta=2.4$ in Ref. \cite{GIE08}, the value of the chiral condensate shows even for $L_5=48$ a rather large deviation by a factor of two from the $L_5 \rightarrow \infty$ limit, whereas even for the low order $N=80$ the polynomial approximation deviates only by a factor of 1.5 from the $N \rightarrow \infty$ limit. Therefore an overlap operator defined from a polynomial approximation including $O(100)$ terms seems to be competitive and in the same regime that could be possibly explored today with Domain-Wall fermions.

\section{Witten index and fermion-boson degeneracy}\label{sec:witten}

Periodic boundary conditions applied to all fields imply that the partition function being simulated is the Witten index, corresponding in the Hamiltonian formalism to the super-trace 
\begin{equation}
W(V_3,L) = \textrm{Tr}\left\{(-1)^F \exp\left( - L\hat{H}(V_3) \right)\right\}\,,
\end{equation}
being $V_3$ the three dimensional volume of the lattice box and $L$ its temporal extent. The Witten index, expressed in a basis of eigenstates of the Hamiltonian $\hat{H}(V_3)$, reads 
\begin{equation}\label{witten}
W(V_3,L) = \sum_{n \in {\textrm{boson}}} \exp\left( - L E_n(V_3) \right) - \sum_{n \in {\textrm{fermions}}} \exp\left( - L E_n(V_3) \right)\,,
\end{equation}
and it is equal to a constant independent of the volume $V_3$ and the temporal extent $L$ if supersymmetry is not broken explicitly, thanks to the cancellation and the matching between boson and fermion states. The energies of each eigenstate $E_n(V_3)$ can depend on the spatial volume, however the Witten index is effectively a topological invariant quantity counting the difference of the boson and fermion number of zero energy vacuum states \cite{WIT82}. 

The idea is to use the Witten index as a probe for supersymmetry restoration on the lattice. Any deviation from a constant behavior of $W(V_3,L)$ is a signal induced by lattice artefacts or by the polynomial approximation. However, the Witten index cannot be easily computed by Monte-Carlo simulations, as it requires knowledge of the absolute normalization of the partition function. An alternative observable could be the vacuum energy density that is the order parameter of (spontaneous) supersymmetry breaking, but the mixing with identity operator that occurs on the lattice due to the breaking of Lorentz symmetry prevents any straight forward implementation. The vacuum energy density, together with the other components of the stress energy momentum tensor, can be computed only indirectly from lattice simulations by studying for instance short distance behavior of suitable correlation functions. 

Fortunately, the ultraviolet divergence coming from the mixing with the identity operator does not depend on $V_3$ nor on $L$. We can therefore study how the Witten index reacts to a simultaneous size change for all directions of the four-dimensional box
\begin{equation}
\Delta \equiv \frac{1}{L^4} \frac{\partial \log W(L^3,L)}{\partial \log L}
\end{equation} 
relative to its infinite volume value. If supersymmetry is realized on the lattice, this derivative of the Witten index must vanish. As the size of the box $L$ is squeezed, more and more excited states contribute to the sum in Eq.~\ref{witten}, until the point where energies close to the ultraviolet lattice cut-off are reached and the approximate degeneracy between fermion and boson states is broken. We expect the size $L$ of the box where this breakdown occurs to be small if our $O(a)$ improved lattice action is able to keep the magnitude of lattice artefacts under control at our current lattice spacings.

\begin{figure}
\centering
 \includegraphics[width=.49\textwidth]{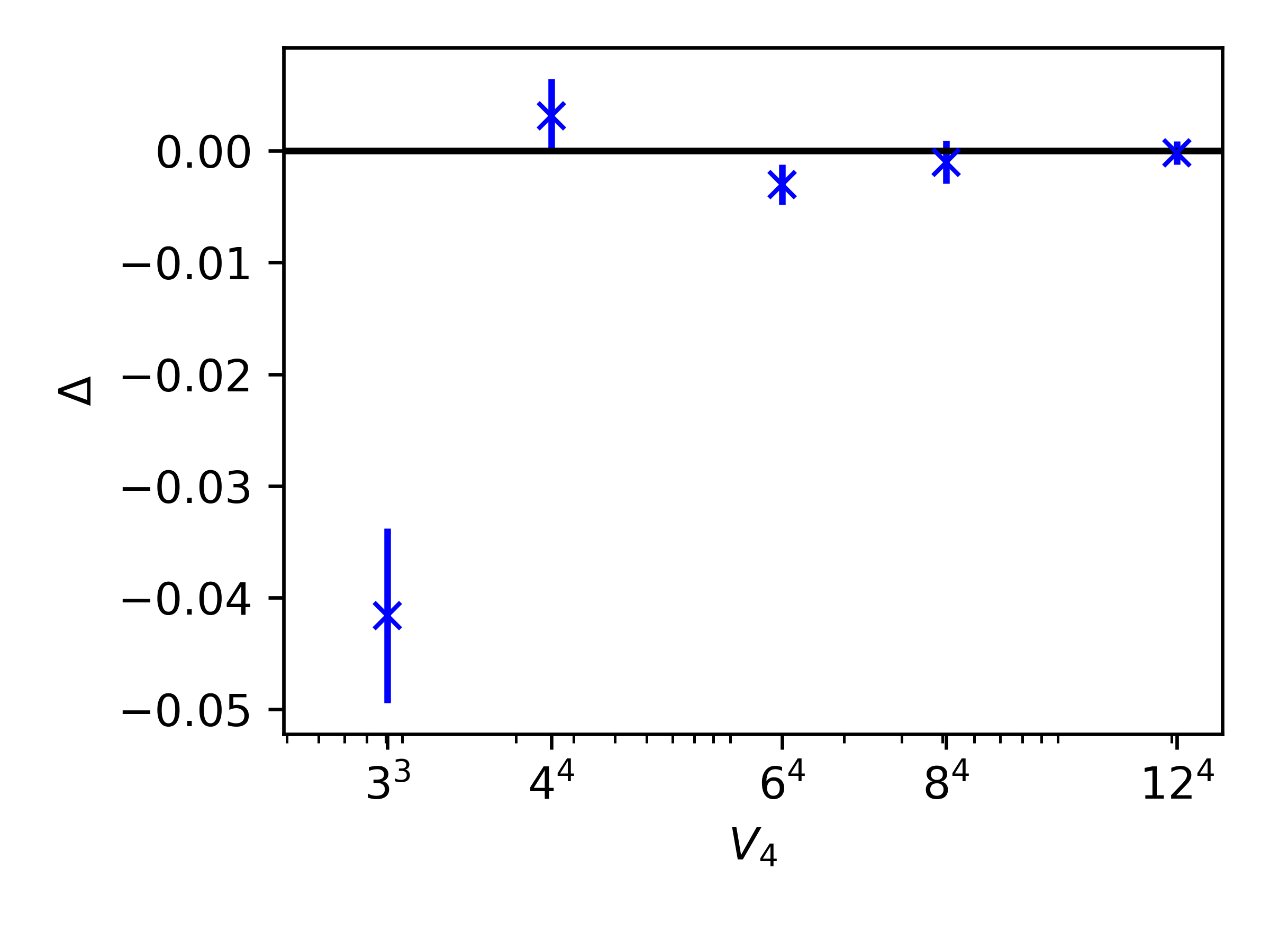}
 \caption{Subtracted derivative of the Witten index $W$ defined as $\Delta \equiv \frac{1}{L^4} \frac{\partial \log W(L^3,L)}{\partial \log L}$ for periodic boundary conditions applied to the gluino field for the simulations with the polynomial approximation $N=250$ for various lattice of volumes $V_4$. Significant deviations from zero of $\Delta$ appear from the smallest volume $V_4=3^4$.}\label{breakdown_witten_index_0000m_1600b_2000k}
\end{figure}

The partial derivative with respect to the box size $L = N a$ can be computed straightforwardly with the chain rule as the derivative of the partition function with respect to the bare gauge coupling
\begin{equation}
\Delta = \frac{4 N_c}{g^3} \frac{\partial g}{\partial \ln(a)}(\langle s_g\rangle - \langle s_g\rangle_0)\,,
\end{equation}
since the lattice spacing $a$ is a function of $g$, $a \equiv a(g)$. The derivative $\Delta$ is proportional to the expectation value of the gauge energy density $\langle s_g\rangle$ minus its infinite-volume vacuum value, up to a multiplicative constant proportional to the $\beta$-function.\footnote{Note that $\Delta$ does not include terms proportional to the chiral condensate, as the bare mass is proportional to the renormalized mass with overlap fermions and it is equal to zero in our simulations.} The dependence of the lattice spacing on the gauge coupling can be estimated using the NSZV expression as
\begin{equation}
\frac{\partial g}{\partial \ln(a)} = -\frac{g^3}{16\pi^2}\frac{3 N_c}{1 - \frac{N_c g^2}{8 \pi^2}}\,,
\end{equation}
that is exact up to one-loop order in perturbation theory in the lattice regularization scheme. As the Witten index is a constant, we expect the quantity $\Delta$ to be zero. Deviations from zero will appear for anti-periodic fermion boundary conditions in time direction as a result of the different statistics of fermion and bosons. In this case $\Delta$ is related to the thermal average of the trace of the energy-momentum tensor, that raises sharply near the deconfinement phase transition. In Fig.~\ref{breakdown_witten_index_0000m_1600b_2000k} the Witten index $\Delta$ is compatible with zero up to a small lattice box of size $4^4$, while the deviation from zero is observed starting from a lattice box $3^4$. We can therefore conclude that overlap fermions are able to keep an effective degeneracy of fermion and boson states on boxes as coarse as $4^4$.

\section{Volume reduction}\label{sec:volume}

Along the same ideas as the Seiberg-Witten electromagnetic duality, several recent theoretical studies opened the possibility for an understanding of confinement in four-dimensional $\mathcal{N}=1$ SYM on $R^4\times S^1$ based on a semiclassical analysis at small compactification radii. The cancellations occurring between boson and fermion states guaranteed by supersymmetry for the Witten index, allows to analytically compute the structure of the phase diagram. $\mathcal{N}=1$ SYM is expected to Abelianize for small compactification radii, with confinement preserved by non-perturbative effects \cite{UNS08,AZE10,UNS10,POP12}. If there are no phase transitions disconnecting the small from the large compactification regimes, semiclassical calculations might be able to provide an analytical understanding of confinement of a four dimensional interacting gauge theory. The first lattice simulations of the compactified $\mathcal{N}=1$ SYM have been able to show evidences for continuity, but the small compactification regime has been difficult to explore given the explicit breaking of chiral symmetry induced by Wilson fermions \cite{BER14,BER18}. Overlap fermions would allow to explore better the small compactification regime, in particular when considering whether chiral symmetry is remains spontaneously broken. 

\begin{figure}
\centering
 \subfigure[Periodic vs antiperiodic]{\includegraphics[width=.49\textwidth]{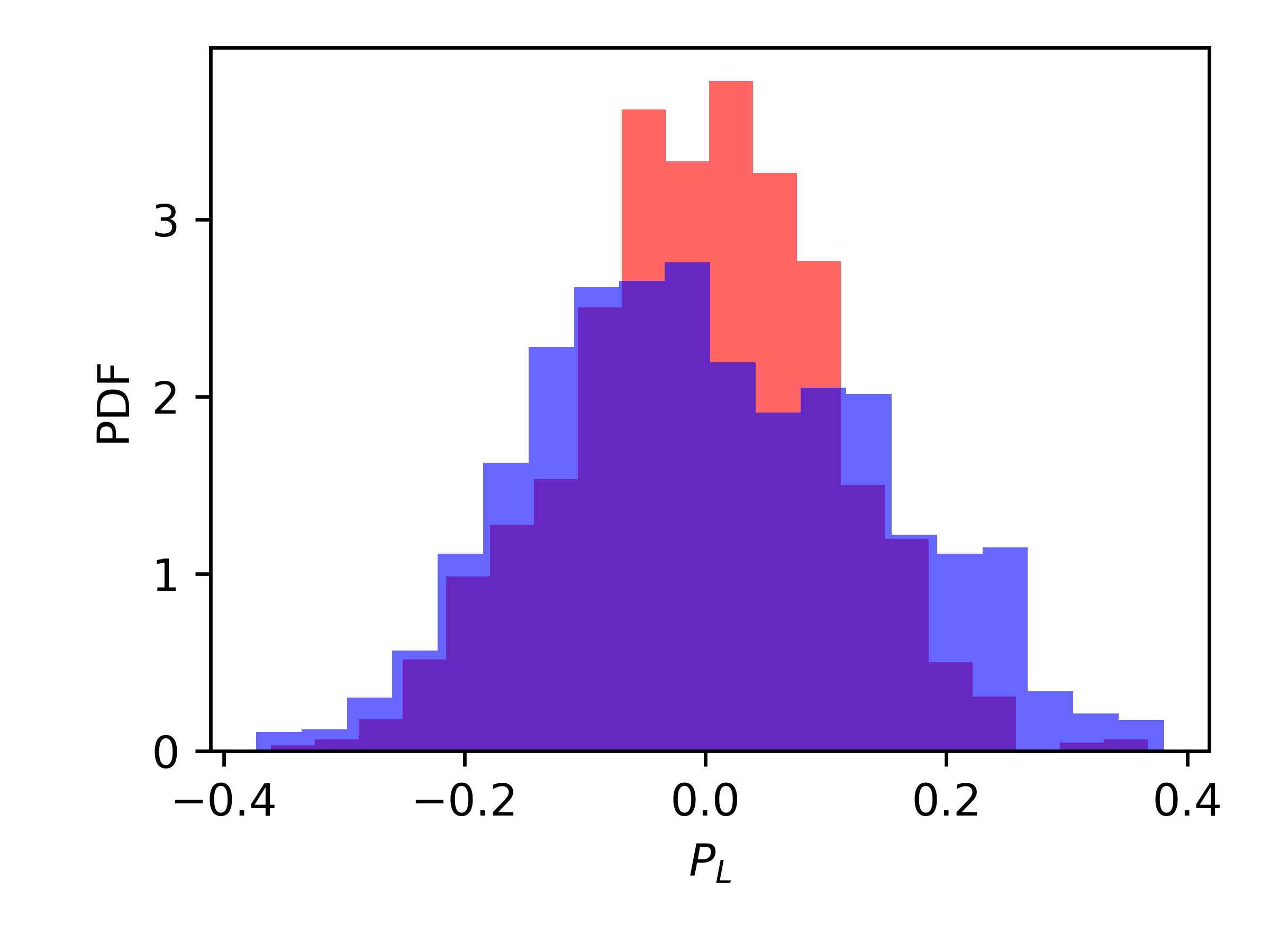}\label{polyakov_loop_periodic_vs_antiperiodic_0000m_1600b_2000k}}
 \subfigure[Volume comparison $4^4$ vs $8^4$]{\includegraphics[width=.49\textwidth]{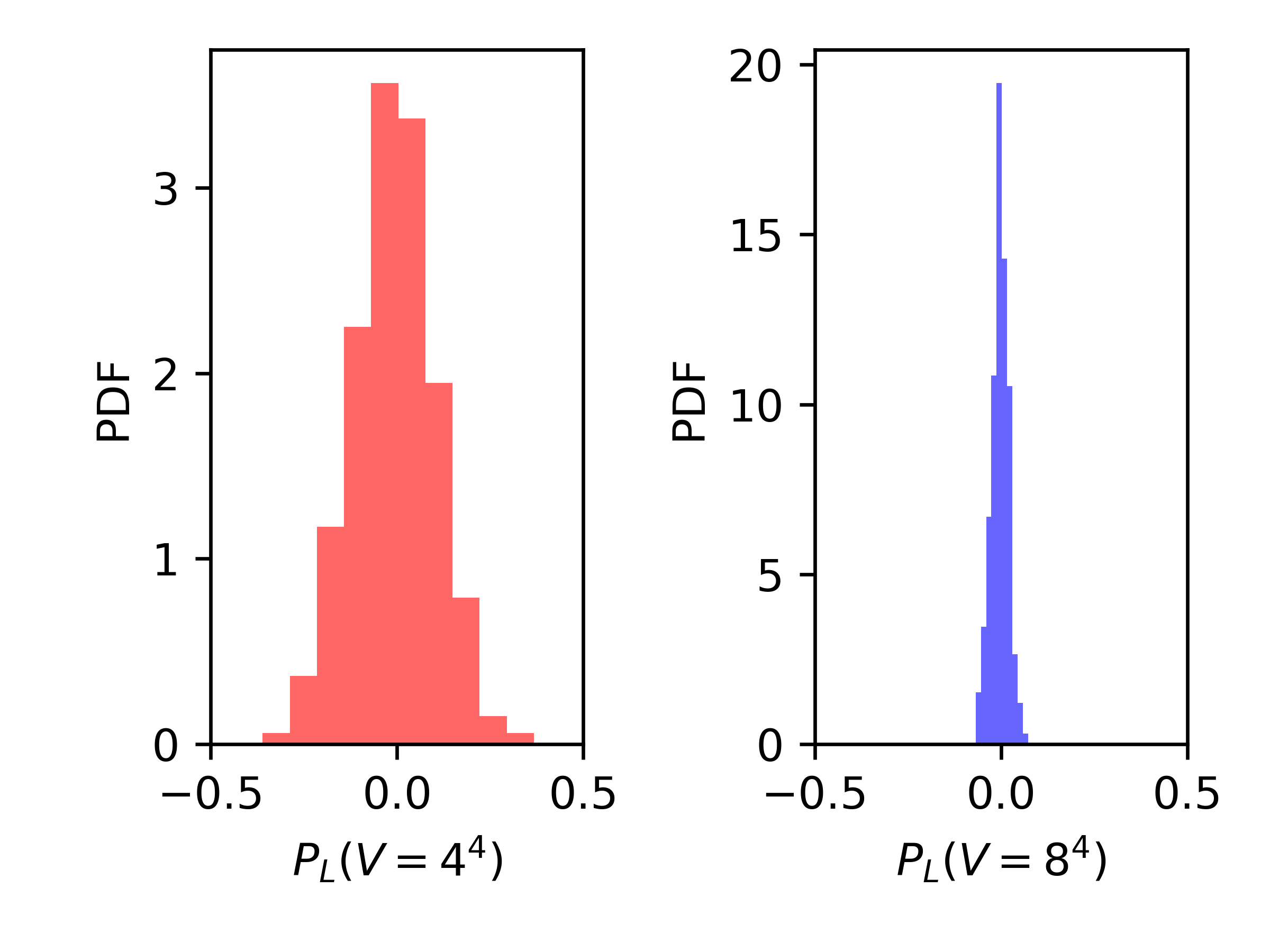}\label{polyakov_loop_4c4_vs_8c8_0000m_1600b_2000k}}
 \caption{a) Distribution of the Polyakov loop for periodic (red) and antiperiodic (blue) on a volume $4^4$ for $N=250$. b) Distribution of the Polyakov loop for periodic boundary conditions on a volume $4^4$ (red) and $8^8$ (blue) for $N=250$.}
\end{figure}

The compactification of more than one space-time dimension has been studied in the context of the Eguchi-Kawai volume reduction \cite{AZE10,UNS10}. In this exploratory study we can provide some evidence for volume independence of bulk quantities such as the gluino condensate, as shown in Fig.~\ref{volume_chiral_condensate_250_0000m_1600b_2000k}. The chiral condensate shows a complete volume independence up to lattices of size $6^4$, while it drops for $4^4$. Given the small number of lattice points of a volume equal to $4^4$, it is difficult to tell if such deviation is a genuine volume dependence of the chiral condensate or rather a lattice artefact. The gauge action, proportional to a combination of square and rectangular Wilson loops, appears to be instead constant even at $4^4$. The flat behavior follows from the Witten index shown in Fig.~\ref{breakdown_witten_index_0000m_1600b_2000k}, which is proportional to the change of the gauge action. The Polyakov loop $P_L$ has a symmetric distribution also for the volume $4^4$ for periodic boundary conditions, while a double peak asymmetric distribution is present for antiperiodic boundary conditions applied to the gluino field, see Fig.~\ref{polyakov_loop_periodic_vs_antiperiodic_0000m_1600b_2000k}. The distribution of $P_L$ becomes broader as the volume gets smaller and smaller, see Fig.~\ref{polyakov_loop_4c4_vs_8c8_0000m_1600b_2000k}, in agreement with the predicted broadening of the effective potential of the Polyakov loop for small volumes.

\section{A first look to the bound state spectrum}\label{sec:spectrum}

\begin{figure}
\centering
\subfigure[$\frac{1}{2}^+$ correlator]{\includegraphics[width=.47\textwidth]{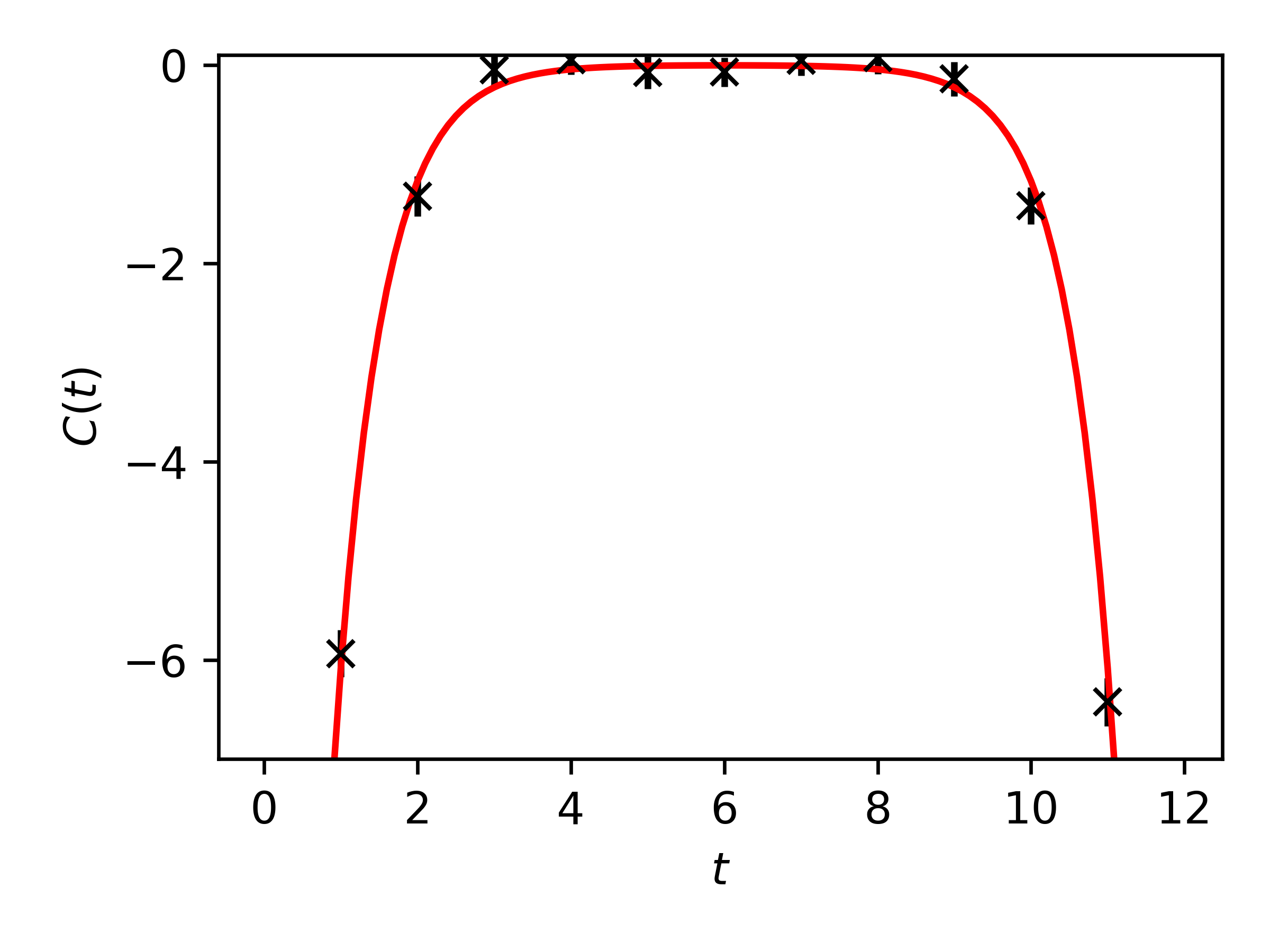}\label{gluinoglue_correlator_0000m_1600b_2000k}}
 \subfigure[Gluino-glue mass as a function of $N$]{\includegraphics[width=.47\textwidth]{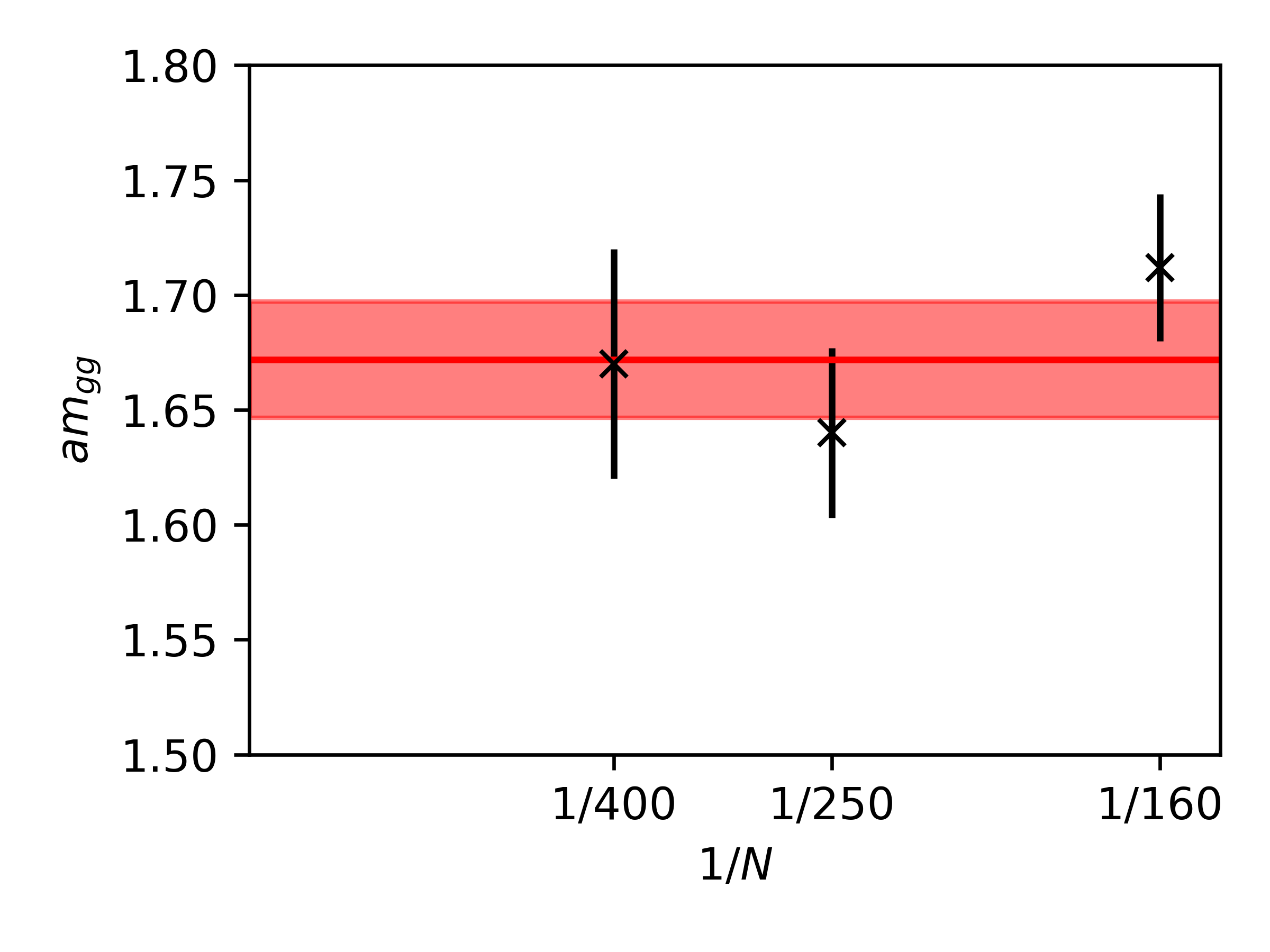}\label{gluinoglue_mass_0000m_1600b_2000k}}
 \caption{a) Correlator of the gluino-glue on the lattice volume $12^4$ and the order of the approximation of the sign function equal to $N=250$. b) Mass of the gluino-glue as a function of the order of the polynomial approximation for the volume $12^4$. The red line and the shaded are represent the central value and the error of the fit to a constant.}
\end{figure}

The bound states of $\mathcal{N}=1$ SYM are expected to be organized in supermultiplets of particles with equal mass \cite{VEN82,FAR97,FAR98,BER1}. The predicted lowest chiral supermultiplet consists of a scalar ($0^+$) and pseudoscalar ($0^-$) bosonic states, that can be identified with a mixture of glueballs and flavor-singlet mesons, and by an exotic spin-$\textrm{1}/{2}$ fermion particle called gluino-glue, typical of theories with fermions in the adjoint representation but without any correspondent in QCD.

The gluino-glue mass is measured from the exponential decay at large time distance of the zero momentum projected correlator of the operator $F^{\mu\nu}(x) \sigma_{\mu\nu} \lambda(x)$. Twelve levels of APE-smearing with smearing parameter $\alpha=0.15$ are applied to the link fields used to construct the clover version of $F^{\mu\nu}(x)$. The gluino-glue correlator is presented in Fig.~\ref{gluinoglue_correlator_0000m_1600b_2000k}. The fitted gluino mass is independent of the polynomial approximation used in the sign function, as shown in Fig.~\ref{gluinoglue_mass_0000m_1600b_2000k}. The fitted chiral value of the gluino mass in units of $w_0/a$ is $w_0 m_{gg} = 1.348(20)$, which is larger than the continuum limit extrapolated value $w_0 m_{gg}=0.93(6)$ using unimproved Wilson fermions \cite{ALI19}, most likely due to lattice artefacts at our current lattice spacings.

\begin{figure}
\centering
\includegraphics[width=.47\textwidth]{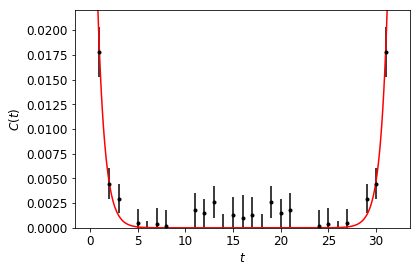}
 \caption{Correlator of the $0^{+}$ glueball on the volume $16^3\times 32$ and the order of the approximation of the sign function equal to $N=160$.}\label{zpp_correlator_160m_160f_16c32_0000m_1600b_2000k}
\end{figure}

The signal in the $0^{++}$ sector suffers from large statistical errors. The relevant $0^{+}$ operators are the fermion bilinear $\bar{\lambda}\lambda$ and the glueball-like operator $F^{\mu\nu}F_{\mu\nu}$. From our previous investigations with Wilson fermions, we have observed that the largest overlap with the ground state is given by glueball-like operators, and therefore we consider a large variational basis of $1\times1$ and $2\times2$ plaquette operators constructed from links with various levels of APE-smearing with smearing parameter $\alpha=0.5$. The measure of the glueball mass has been possible only on the $16^3 \times 32$ lattice with the order of the approximation of the sign function equal to $N=160$, where the fitted mass is $a M_{0^+} = 1.40(53)$, see Fig.~\ref{zpp_correlator_160m_160f_16c32_0000m_1600b_2000k}, a value that is compatible within the large errors with the gluino-glue mass.

\begin{figure}
\centering
\includegraphics[width=.47\textwidth]{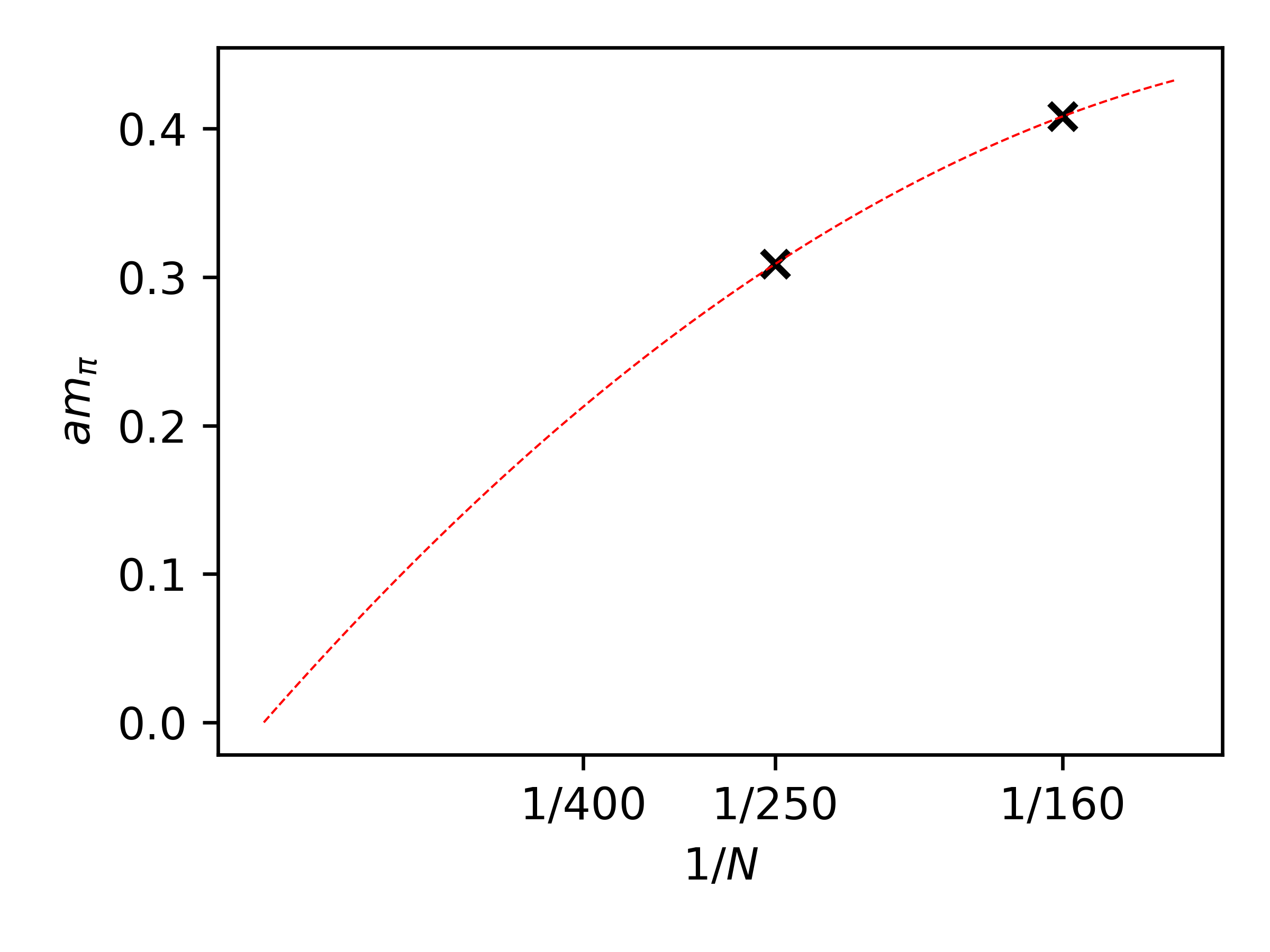}
 \caption{Pion mass as a function of $1/N$ for the volume $16^3\times32$. The red-dashed line is a quadratic interpolation of the two available data with a constant term equal to zero.}\label{pion_mass_0000m_1600b_2000k}
\end{figure}

The adjoint pion is not a physical particle of the theory, but it is defined in the context of partially quenched chiral perturbation theory. 
Its mass can be extracted from the connected part of the correlator of the $\eta'$ particle defined by the operator $\bar{\lambda}\gamma_5\lambda$. 
In the chiral limit, the pion mass is expected to vanish and the $\eta'$ mass is dominated by the disconnected parts of the $0^{-}$ correlators. The square of the pion mass is, according to the approach of partially quenched chiral perturbation theory, proportional to the gluino mass. In our overlap simulations, the gluino mass is set to zero, but the finite order of the approximation leads (similar to the finite $L_5$ for Domain-Wall fermions) to a finite effective mass. It can be measured by the partially conserved axial current relation or the square of the adjoint pion mass. The zero mass limit is obtained in the limit of infinite $N$ without any need for fine-tuning. The pion mass was measured only on the $16^3\times32$ ensembles, where we have a sufficiently large temporal extent for the exponential fit, see Fig.~\ref{pion_mass_0000m_1600b_2000k}. As expected, the pion mass is decreasing from $N=160$ to $N=250$.

\section{Outlook}
Based on this first exploratory studies, Ginsparg-Wilson fermions provide a starting point for several interesting further investigations of supersymmetric theories. In the following, we list a few directions which we are considering for our next studies.

The index theorem relates the difference of the zero eigenvalues with positive and negative chirality of the mass-less overlap operator to the topological charge \cite{LUS99}. Since gluinos are Majorana fermions transforming in the adjoint representation of SU(2), the number of real eigenvalues of the Dirac-Wilson operator is a multiple of four. There is the possibility for a configuration, depending on the boundary conditions, to have for instance just a couple of isolated negative eigenvalues so that Index$(D_{\textrm{adjoint}}) \neq 2 N_c Q_{\textrm{top}}$. Such a configuration would correspond to a ``fractional topological charge'' instanton, semiclassical objects that might be responsible for the gluino condensation \cite{COH84,FOD09,LEU92}. The investigation of fractional topological charges has not been addressed in unquenched simulations, and it would be a relevant application for future simulations with overlap gluinos.

If two dimensions are compactfied, and if their length is reduced to zero, the lower dimensional effective theory has an enlarged supersymmetry. The two-dimensional model would be an ideal benchmark to study the renormalization properties of theories with extended supersymmetry constructed from $\mathcal{N}=1$ SYM with overlap fermions, to connect our simulations to previous results \cite{SUZ05,AUG16,AUG17,AUG18}. 

The improvements guaranteed by overlap fermions in supersymmetric Yang-Mills theory with more than one conserved supercharge or coupled to matter chiral superfields have not been fully addressed so far. Several difficulties appeared even in the simpler case of the Wess-Zumino model, in particular for the consistency between the Majorana condition and chiral symmetry in the Yukawa couplings \cite{FUJ02}. Further difficulties arise from the large number of couplings to be tuned. Even if the fermion mass is protected by chiral symmetry, the bare scalar mass must be tuned such that its renormalized value is zero and there are the Yukawa and the quartic couplings that must be tuned to be equal to the gauge coupling. A large number of simulations would be required only to extrapolate to the supersymmetric limit. Some preliminary study about this possibility in the context of SuperQCD have been presented in Ref.~\cite{BER18b,WEL18}.

\section{Conclusions}

The main advantage of preserving exact chiral symmetry is that $\mathcal{N}=1$ Super Yang-Mills can be simulated on the lattice without the need of fine-tuning any parameter. Further, the scaling of physical observables with overlap fermions to the continuum limit is automatically $O(a)$ improved. The exploratory study of overlap gluino simulations presented in this contribution is promising, as the polynomial approximation of the sign function is able to regularize mass-less chiral fermions, while avoiding at the same time topological freezing and the sign problem. From this perspective, our setup is close to the Domain-Wall implementation of the Ginsparg-Wilson relation, with the role of the size of the fifth dimension replaced by the order $N$ of the polynomial approximation. 

The infrared observables we have studied in the current paper are either independent on the order $N$ of the polynomial approximation, such as the gluino-glue mass or the scale $w_0/a$, or they show a mild linear dependence, as in the case of the gluino condensate. Our results are in agreement with the spontaneous breaking of chiral symmetry, as the gluino condensate is clearly non-vanishing in the chiral limit $N \rightarrow \infty$. The bare value of the gluino condensate in dimensionless units is close to the value determined by previous investigations of the same theory using Domain-Wall fermions or using Wilson fermions with the gradient flow, confirming their agreement up to an $O(1)$ renormalization constant.

The Witten index, and in particular its derivative, allows to check and confirm the expected boson-degeneracy required by unbroken supersymmetry. Our numerical results are compatible with a constant Witten index up to lattice boxes as small as $4^4$ at the current lattice spacing. As glueball states are affected by a bad signal-to-noise ratio, further studies and larger statistics will be required to confirm the expected fermion-boson degeneracy directly from the bound state spectrum.

\section*{Acknowledgments}
We thank G.~M\"unster and I.~Montvay for helpful comments and discussions.
The authors gratefully acknowledge the Gauss Centre for Supercomputing e.V. (www.gauss-centre.eu) for funding this project by providing computing time on the GCS Supercomputer SuperMUC at Leibniz Supercomputing Centre (www.lrz.de).
Further computing time has been provided on the compute cluster PALMA
of the University of M\"unster. G.~Bergner and C.~Lopez acknowledge support from the Deutsche Forschungsgemeinschaft (DFG) Grant
No.~BE 5942/2-1.

\appendix
\section{Summary table of analyzed runs}\label{sec:summary}

\begin{center}
\begin{tabular}{c|c|c|c}
$L_s$ & $L_t$ & $N$ & number of thermalized \\
 &  &  & configurations \\
 \hline
12       & 12      &   80 & 1000   \\
8         & 8        & 160 & 1000 \\
12       & 12      & 160 & 700   \\
16       & 32      & 160 & 700   \\
3         & 3        & 250 & 2500 \\
4         & 4        & 250 & 2500 \\
6         & 6        & 250 & 1000  \\
8         & 8        & 250 & 700  \\
12       & 12      & 250 & 700  \\
16       & 32      & 250 & 300  \\
8         & 8        & 400 & 500  \\
12       & 12      & 400 & 400  \\
\end{tabular}
\end{center}

\end{document}